\documentclass[10pt, letterpaper, twocolumn]{IEEEtran}

\usepackage[mathscr]{eucal}
\usepackage[cmex10]{amsmath}
\usepackage{epsfig,epsf,psfrag}
\usepackage{amssymb,amsmath,amsthm,amsfonts,latexsym}
\usepackage{amsmath,graphicx,bm,xcolor,url}
\usepackage[caption=false]{subfig} 
\usepackage{fixltx2e}
\usepackage{array}
\usepackage{verbatim}
\usepackage{bm}
\usepackage{algorithmic, cite}
\usepackage{algorithm}
\usepackage{verbatim}
\usepackage{textcomp}
\usepackage{mathrsfs}
\usepackage{epstopdf}

\catcode`~=11 \def\UrlSpecials{\do\~{\kern -.15em\lower .7ex\hbox{~}\kern .04em}} \catcode`~=13 

\allowdisplaybreaks[4]
 
\newcommand{\nn}{\nonumber}

\newcommand{\calA}{\mathcal{A}}
\newcommand{\calB}{\mathcal{B}}
\newcommand{\calC}{\mathcal{C}}
\newcommand{\calD}{\mathcal{D}}
\newcommand{\calE}{\mathcal{E}}
\newcommand{\calF}{\mathcal{F}}

\newcommand{\calI}{\mathcal{I}}
\newcommand{\calJ}{\mathcal{J}}
\newcommand{\calK}{\mathcal{K}}
\newcommand{\calL}{\mathcal{L}}
\newcommand{\calM}{\mathcal{M}}

\newcommand{\calO}{\mathcal{O}}
\newcommand{\calP}{\mathcal{P}}

\newcommand{\calS}{\mathcal{S}}
\newcommand{\calT}{\mathcal{T}}
\newcommand{\calU}{\mathcal{U}}
\newcommand{\calV}{\mathcal{V}}
\newcommand{\calW}{\mathcal{W}}
\newcommand{\calX}{\mathcal{X}}
\newcommand{\calY}{\mathcal{Y}}

\newcommand{\bS}{\mathbf{S}}

\newcommand{\bU}{\mathbf{U}}

\newcommand{\bV}{\mathbf{V}}

\newcommand{\bW}{\mathbf{W}}

\newcommand{\bX}{\mathbf{X}}

\newcommand{\bY}{\mathbf{Y}}

\newcommand{\bZ}{\mathbf{Z}}

\newcommand{\rma}{\mathrm{a}}

\newcommand{\rmb}{\mathrm{b}}

\newcommand{\rmc}{\mathrm{c}}

\newcommand{\bbE}{\mathbb{E}}

\newcommand{\bbN}{\mathbb{N}}

\newcommand{\bbP}{\mathbb{P}}

\newcommand{\bbR}{\mathbb{R}}

\newcommand{\scC}{\mathscr{C}}

\DeclareMathAlphabet{\mathbsf}{OT1}{cmss}{bx}{n}
\DeclareMathAlphabet{\mathssf}{OT1}{cmss}{m}{sl}

\newcommand{\rvc}{\mathsf{c}}

\newcommand{\rvi}{\mathsf{i}}

\DeclareSymbolFont{bsfletters}{OT1}{cmss}{bx}{n}  
\DeclareSymbolFont{ssfletters}{OT1}{cmss}{m}{n}
\DeclareMathSymbol{\bsfGamma}{0}{bsfletters}{'000}
\DeclareMathSymbol{\ssfGamma}{0}{ssfletters}{'000}
\DeclareMathSymbol{\bsfDelta}{0}{bsfletters}{'001}
\DeclareMathSymbol{\ssfDelta}{0}{ssfletters}{'001}
\DeclareMathSymbol{\bsfTheta}{0}{bsfletters}{'002}
\DeclareMathSymbol{\ssfTheta}{0}{ssfletters}{'002}
\DeclareMathSymbol{\bsfLambda}{0}{bsfletters}{'003}
\DeclareMathSymbol{\ssfLambda}{0}{ssfletters}{'003}
\DeclareMathSymbol{\bsfXi}{0}{bsfletters}{'004}
\DeclareMathSymbol{\ssfXi}{0}{ssfletters}{'004}
\DeclareMathSymbol{\bsfPi}{0}{bsfletters}{'005}
\DeclareMathSymbol{\ssfPi}{0}{ssfletters}{'005}
\DeclareMathSymbol{\bsfSigma}{0}{bsfletters}{'006}
\DeclareMathSymbol{\ssfSigma}{0}{ssfletters}{'006}
\DeclareMathSymbol{\bsfUpsilon}{0}{bsfletters}{'007}
\DeclareMathSymbol{\ssfUpsilon}{0}{ssfletters}{'007}
\DeclareMathSymbol{\bsfPhi}{0}{bsfletters}{'010}
\DeclareMathSymbol{\ssfPhi}{0}{ssfletters}{'010}
\DeclareMathSymbol{\bsfPsi}{0}{bsfletters}{'011}
\DeclareMathSymbol{\ssfPsi}{0}{ssfletters}{'011}
\DeclareMathSymbol{\bsfOmega}{0}{bsfletters}{'012}
\DeclareMathSymbol{\ssfOmega}{0}{ssfletters}{'012}

\newcommand{\hatl}{\hat{l}}

\newcommand{\till}{\tilde{l}}

\newcommand{\hatm}{\hat{m}}
\newcommand{\hatM}{\hat{M}}

\newcommand{\tilR}{\tilde{R}}

\newcommand{\hatS}{\hat{S}}

\newcommand{\hatU}{\hat{U}}

\newcommand{\hatV}{\hat{V}}

\newcommand{\tilV}{\tilde{V}}

\newcommand{\tilW}{\tilde{W}}

\newcommand{\hatX}{\hat{X}}

\newcommand{\tilX}{\tilde{X}}

\newcommand{\hatY}{\hat{Y}}

\newcommand{\veps}{\varepsilon}

\def\fndot{\, \cdot \,}

\newcommand{\bone}{\mathbf{1}}

 \def\independenT#1#2{\mathrel{\rlap{$#1#2$}\mkern5mu{#1#2}}}
\newcommand\indep{\protect\mathpalette{\protect\independenT}{\perp}}

\newtheorem{theorem}{Theorem} 
\newtheorem{lemma}[theorem]{Lemma}

\newtheorem{corollary}[theorem]{Corollary}
\newtheorem{definition}{Definition} 
\newtheorem{example}{Example} 
 
\newtheorem{remark}{Remark}

\newcommand{\qednew}{\nobreak \ifvmode \relax \else
      \ifdim\lastskip<1.5em \hskip-\lastskip
      \hskip1.5em plus0em minus0.5em \fi \nobreak
      \vrule height0.75em width0.5em depth0.25em\fi}

\DeclareMathOperator*{\plimsup}{\mathfrak{p}-lim\, sup\,}
\DeclareMathOperator*{\pliminf}{\mathfrak{p}-lim\, inf\,}
\newcommand{\underI}{\underline{I}}
\newcommand{\overI}{\overline{I}}
\newcommand{\underH}{\underline{H}}
\newcommand{\overH}{\overline{H}} 
\newcommand{\Rd}{R_{\mathrm{d}}}

\newcommand{\frakc}{\rvc}

\allowdisplaybreaks[2]

\title{A Formula for the Capacity of the General Gel'fand-Pinsker Channel}
\author{Vincent Y.~F.\ Tan, {\em Member, IEEE} \thanks{The author is with the Department  of Electrical and Computer Engineering as well as the Department  of  Mathematics,  National University of Singapore (email:  vtan@nus.edu.sg). This paper was presented in part at the 2013 International Symposium on Information Theory in Istanbul, Turkey.}}

\begin{document}
\flushbottom
\maketitle

\begin{abstract}
We consider  the Gel'fand-Pinsker problem in which the channel and state are general, i.e.,   possibly   non-stationary,  non-memoryless and non-ergodic. Using  the information spectrum method and a non-trivial modification of   the   piggyback coding lemma by  Wyner, we prove that the capacity can be expressed as an  optimization over the difference of  a spectral  inf-   and  a spectral  sup-mutual information rate.  We  consider various specializations including the case where the channel and state are memoryless but not necessarily  stationary. 
\end{abstract}


\section{Introduction}\label{sec:intro}
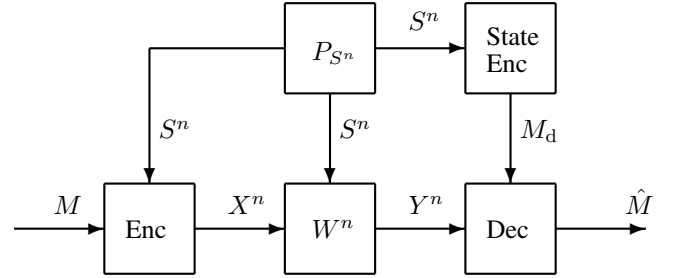
\begin{figure}
\centering
\setlength{\unitlength}{.4mm}
\begin{picture}(200, 90)
\thicklines
\put(0, 15){\vector(1, 0){30}}
\put(60, 15){\vector(1,0){30}}
\put(120, 15){\vector(1,0){30}}
\put(180, 15){\vector(1,0){30}}
\put(30, 0){\line(1, 0){30}}
\put(30, 0){\line(0,1){30}}
\put(60, 0){\line(0,1){30}}
\put(30, 30){\line(1,0){30}}

\put(90, 0){\line(1, 0){30}}
\put(90, 0){\line(0,1){30}}
\put(120, 0){\line(0,1){30}}
\put(90, 30){\line(1,0){30}}

\put(10, 20){  $M$}
\put(68, 20){  $X^n$}
\put(128, 20){  $Y^n$} 
\put(37, 12){Enc} 
\put(99, 12){$W^n$} 

\put(150, 0){\line(1, 0){30}}
\put(150, 0){\line(0,1){30}}
\put(180, 0){\line(0,1){30}}
\put(150, 30){\line(1,0){30}}
\put(157, 12){Dec} 
\put(200, 20){  $\hatM $} 

\put(90, 60){\line(1, 0){30}}
\put(90, 60){\line(0,1){30}}
\put(120, 60){\line(0,1){30}}
\put(90, 90){\line(1,0){30}}
\put(120, 75){\vector(1,0){30}}
\put(165, 60){\vector(0,-1){30}}
\put(105, 60){\vector(0,-1){30}}
\put(90, 75){\line(-1,0){45}}
\put(45, 75){\vector(0,-1){45}}

\put(150, 60){\line(1, 0){30}}
\put(150, 60){\line(0,1){30}}
\put(180, 60){\line(0,1){30}}
\put(150, 90){\line(1,0){30}}

\put(105, 45){  $S^n$} 
\put(45, 45){  $S^n$} 
\put(165, 45){  $M_{\mathrm{d}}$} 
\put(128, 81){  $S^n$} 
\put(99, 71){$P_{S^n}$} 
\put(157, 76){State} 
\put(157, 67){Enc} 
  \end{picture}
  \caption{The Gel'fand-Pinsker problem with rate-limited coded state information available at the decoder~\cite{heegard, Ste08}. The message $M$ is uniformly distributed in $[1:\exp(nR)]$ and independent of $S^n$. The compression index $M_{\mathrm{d}}$ is rate-limited and takes  values in $[1:\exp(n\Rd)]$.  The canonical Gel'fand-Pinsker problem~\cite{GP80} is a special case in which the output of the state encoder is a deterministic quantity. }
  \label{fig:gp}
\end{figure}

In this paper,  we consider the classical problem of channel coding with noncausal state information at the encoder, also know as the {\em Gel'fand-Pinsker} problem~\cite{GP80}. In this problem, we would like to send a uniformly distributed message  over a state-dependent channel $W^n:\calX^n\times\calS^n\to\calY^n$, where $\calS,\calX$ and $\calY$ are the state, input and output alphabets respectively. The random state sequence  $S^n\sim P_{S^n}$ is available noncausally at the encoder but not at the decoder. See Fig.~\ref{fig:gp}. The Gel'fand-Pinsker  problem   consists in finding the maximum rate for which there exists a reliable code. Assuming that the channel and state sequence are stationary and memoryless, Gel'fand and Pinsker~\cite{GP80} showed that this maximum message rate or {\em capacity} $C=C(W,P_S)$ is given by 
\begin{equation}
C=\max_{\substack{P_{U|S}, \, g:\calU\times\calS\to\calX\\|\calU|\le|\calX| |\calS|+1}}I(U;Y)-I(U;S). \label{eqn:gp_capacity}
\end{equation}
The coding scheme involves a covering step at the encoder to reduce the uncertainty due to the random state sequence and a packing step  to decode the message~\cite[Chapter~7]{elgamal}.  Thus, we observe  the covering rate  $I(U;S)$ and the packing rate $I(U;Y)$ in~\eqref{eqn:gp_capacity}. A weak converse can be  proved by using the Csisz\'{a}r-sum-identity~\cite[Chapter~7]{elgamal}. A strong converse was proved by Tyagi and Narayan~\cite{tyagi} using entropy and image-size  characterizations~\cite[Chapter~15]{Csi97}.  The Gel'fand-Pinsker problem has numerous applications, particularly in information hiding and watermarking~\cite{Mou03}. 

In this paper, we revisit the Gel'fand-Pinsker problem. Instead of assuming stationarity and memorylessness on the channel and state sequence,  we let the channel $W^n$ be a {\em general} one in the sense of Verd\'{u} and Han~\cite{VH94,Han10}. That is,   $\bW=\{W^n\}_{n=1}^{\infty}$ is an arbitrary sequence of stochastic mappings from $\calX^n\times\calS^n$ to $\calY^n$.  We also model the state distribution as a {\em general} one $\bS\sim\{P_{S^n} \in \calP(\calS^n) \}_{n=1}^{\infty}$.  Such generality  allows us to   understand optimal coding schemes for general systems. We  prove an analogue of the Gel'fand-Pinsker capacity in~\eqref{eqn:gp_capacity} by using information spectrum analysis~\cite{Han10}. Our result  is expressed in terms of the limit superior and limit inferior  in probability operations~\cite{Han10}.  For the direct part, we leverage on a technique  used by Iwata and Muramatsu~\cite{iwata02} for the general Wyner-Ziv problem~\cite[Chapter~11]{elgamal}.   Our proof technique involves a  non-trivial modification of Wyner's piggyback coding  lemma (PBL)~\cite[Lemma~4.3]{Wyner75}. We  also       find the capacity region for the case where {\em rate-limited coded state information} is available at the decoder. This setting, shown in Fig.~\ref{fig:gp}, was studied  by Steinberg~\cite{Ste08} but we consider the scenario in which   the state and   channel are   general.

\subsection{Main Contributions}

There are two main contributions in this work. 

First by developing a non-asymptotic upper bound on the average probability  for any Gel'fand-Pinsker problem (Lemma~\ref{lem:bound_err}), we prove that the general capacity  is 
\begin{equation}
C=\sup \underI(\bU;\bY)-\overI(\bU;\bS),  \label{eqn:general_gp}   
\end{equation}
where the supremum is over all conditional probability laws $\{ P_{X^n, U^n|S^n}\}_{n=1}^{\infty}$ (or equivalently over Markov chains $U^n-(X^n,S^n)-Y^n$ given the channel law $W^n$) and $\underI(\bU;\bY)$ (resp.\  $\overI(\bU;\bS)$) is the spectral inf-mutual information rate (resp.\ the spectral sup-mutual information rate) \cite{Han10}.  The expression in~\eqref{eqn:general_gp} reflects the fact that there are two distinct steps: a covering step and packing step. To cover successfully, we need to expend a rate of $\overI(\bU;\bS)+\gamma$  (for any $\gamma>0$) as stipulated by general fixed-length  rate-distortion theory~\cite[Section~VI]{Ste96}. Thus, each subcodebook has to have at least $\approx\exp( n(\overI(\bU;\bS)+\gamma))$ sequences.  To decode  the codeword's subcodebook index correctly, we can have at most $\approx\exp(n (\underI(\bU;\bY)- \gamma))$  codewords by the general channel coding result of Verd\'{u} and Han~\cite{VH94}. We can  specialize the general result in~\eqref{eqn:general_gp} to the following scenarios: (i) no state information, thus recovering the result by Verd\'{u} and Han~\cite{VH94},   (ii) common state information is available at the encoder and the decoder,   (iii) the channel and state are  memoryless (but not necessarily stationary) and (iv) mixed channels and  states~\cite{Han10}. 

Second, we extend the above result to the case where coded state information is available at the decoder. This problem was first studied by Heegard and El Gamal~\cite{heegard} and later by Steinberg~\cite{Ste08}. In this case, we combine our coding scheme with that of Iwata and Muramatsu for the general Wyner-Ziv problem~\cite{iwata02} to obtain the tradeoff between $\Rd$, the rate of the compressed state information that is available at the decoder, and $R$ be the message rate. We show that the  tradeoff (or capacity region) is the set of rate pairs $(R,\Rd)$ satisfying 
\begin{align}
\Rd&\ge \overI(\bV;\bS)-\underI(\bV;\bY) \label{eqn:wz_gen}\\
R&\le \underI(\bU;\bY|\bV)-\overI(\bU;\bS|\bV),\label{eqn:cgp_gen}
\end{align}
for some  $(\bU,\bV)-(\bX,\bS)-\bY$.   This  general result  can be specialized the stationary, memoryless setting~\cite{Ste08}.

\subsection{Related Work}
The study of general channels started with the seminal work  by Verd\'{u} and Han~\cite{VH94} in which the authors characterized the capacity  in terms of the limit inferior in probability of a sequence of information densities. See Han's book~\cite{Han10} for a comprehensive exposition on the information spectrum method. This line of analysis provides deep insights into the fundamental limits of the transmission of information over general channels and the compressibility of general sources that may not be stationary, memoryless or ergodic. Information spectrum analysis has been used for rate-distortion~\cite{Ste96}, the Wyner-Ziv problem~\cite{iwata02}, the Heegard-Berger problem~\cite{mat12}, the Wyner-Ahlswede-K\"orner (WAK) problem~\cite{miyake}   and the wiretap channel~\cite{Hayashi06, bloch13}. The Wyner-Ziv and wiretap problems are the most closely related to the problem we solve in this paper. In particular, they involve differences of mutual informations akin to the Gel'fand-Pinsker problem.  Even though it is not in the   spirit of information spectrum methods, the  work of Yu {\em et al.}~\cite{WeiYu} deals with the Gel'fand-Pinsker  problem for non-stationary, non-ergodic Gaussian noise and state (also called ``writing on colored paper''). We contrast our work to~\cite{WeiYu}  in Section~\ref{sec:idea}.

It is also worth mentioning that bounds on the reliability function (error exponent) for the Gel'fand-Pinsker problem have been derived by Tyagi and Narayan~\cite{tyagi} (upper bounds)  and  Moulin and Wang~\cite{Mou07} (lower bounds). 

\subsection{Paper Organization}
The rest of this paper is structured as follows. In Section~\ref{sec:sys}, we state our notation and various other definitions. In Section~\ref{sec:info}, we state   all information spectrum-based results and their specializations.   We conclude our discussion in Section~\ref{sec:concl}. The proofs of all results are provided in the Appendices. 
\section{System Model and Main Definitions} \label{sec:sys}
In this section, we state our notation and the   definitions of the various problems that we  consider in this paper.
\subsection{Notation}
Random variables (e.g., $X$) and their realizations (e.g., $x$)  are denoted by upper case and lower case serif font respectively. Sets are denoted in calligraphic font (e.g., the alphabet of $X$ is $\calX$). We use the notation $X^n$ to mean a vector of random variables $(X_1,\ldots, X_n)$. In addition, $\bX=\{X^n\}_{n=1}^{\infty}$ denotes a general source in the sense that each member of the sequence $X^n = (X^{(n)}_1,\ldots, X^{(n)}_n)$ is a random vector. The same holds for a general channel $\bW=\{W^n:\calX^n\to\calY^n\}_{n=1}^{\infty}$, which is simply a sequence of stochastic mappings from $\calX^n$ to $ \calY^n$. The set of all probability distributions with support on an alphabet $\calX$ is denoted as $\calP(\calX)$. We use the notation $X\sim P_X$ to mean that the distribution of $X$ is $P_X$.  The joint distribution induced by the marginal $P_X$ and the conditional $P_{Y|X}$ is denoted as $P_{X}\circ P_{Y|X}$. Information-theoretic quantities are denoted using the usual notations~\cite{Han10, Csi97} (e.g., $I(X;Y)$ for mutual information or $I(P_X, P_{Y|X})$ if we want to make the dependence on distributions explicit).   All logarithms are to the base $\mathrm{e}$. We also use the discrete interval \cite{elgamal} notation $[i:j]:=\{i,\ldots,j\}$  and, for the most part, ignore integer requirements. 

We recall the following probabilistic limit operations~\cite{Han10}.
\begin{definition}
Let $\bU:=\{U_n\}_{n=1}^{\infty}$ be a sequence of real-valued random variables.  The {\em limsup in probability} of $\bU$ is an extended real-number defined as 
\begin{equation}
\plimsup_{n\to\infty} U_n:= \inf\left\{\alpha :  \lim_{n\to\infty}\bbP(U_n>\alpha)=0\right\}.
\end{equation}
The {\em liminf in probability} of $\bU$ is defined as  
\begin{equation}
\pliminf_{n\to\infty} U_n :=-\plimsup_{n\to\infty} ( -U_n).
\end{equation}
\end{definition}
 We also recall the following definitions from Han~\cite{Han10}. These definitions  play a prominent role in the following.
\begin{definition}
Given a pair of   stochastic processes $( \bX,\bY)=\{ X^n,Y^n\}_{n=1}^{\infty}$ with joint distributions $\{P_{X^n, Y^n}\}_{n=1}^{\infty}$, the {\em spectral sup-mutual information rate} is defined as 
\begin{equation}
\overI(\bX;\bY):=\plimsup_{n\to\infty}\frac{1}{n}\log \frac{P_{Y^n|X^n}(Y^n|X^n)}{P_{Y^n}(Y^n)}. \label{eqn:plimsup}
\end{equation}
The {\em spectral  inf-mutual information rate} $\underI(\bX;\bY)$  is defined as in~\eqref{eqn:plimsup} with $\pliminf$ in place of $\plimsup$. The spectral sup- and inf-conditional mutual information rates  are defined similarly. 
\end{definition}
\subsection{The Gel'fand-Pinsker Problem}
We now recall the definition of the Gel'fand-Pinsker problem. The setup is shown in Fig.~\ref{fig:gp} with $M_{\mathrm{d}}=\emptyset$. 
\begin{definition} \label{def:gp_prob}
An $(n, M_n, \epsilon)$ {\em  code for the Gel'fand-Pinsker problem} with channel $W^n:\calX^n\times\calS^n\to\calY^n$ and state distribution $P_{S^n} \in\calP(\calS^n)$ consists of 
\begin{itemize}
\item An  encoder $f_n: [1:M_n]\times\calS^n\to\calX^n$ (possibly randomized)
\item A decoder $\varphi_n:\calY^n\to [1:M_n]$ 
\end{itemize}
such that the average error probability in decoding the message  does not exceed $\epsilon$, i.e.,
\begin{equation}
 \frac{1}{M_n}\sum_{s^n\in\calS^n}P_{S^n}(s^n)\sum_{m=1}^{M_n} W^n( \calB_m^c | f_n(m,s^n),s^n)\le\epsilon,  \label{eqn:error_av}
\end{equation}
where $\calB_m:=\{y^n\in\calY^n:\varphi_n(y^n)=m\}$ and $\calB_m^c:=\calY^n\setminus\calB_m$. 
\end{definition}
We assume that the message is uniformly distributed in the message set $[1:M_n]$ and that it is independent of the state sequence $S^n\sim P_{S^n}$.  Here we remark that in~\eqref{eqn:error_av} (and everywhere else in the paper), we use the notation $\sum_{s^n \in\calS^n}$ even though $\calS^n$ may not be countable. This is the convention we adopt throughout the paper even though integrating against the measure $P_{S^n}$ as in $\int_{\calS^n} \fndot \mathrm{d}P_{S^n}$ would be  more precise. 
\begin{definition} \label{def:gp}
We say that the nonnegative number $R$  is an {\em $\epsilon$-achievable rate} if there exists a sequence of $(n,M_n,\epsilon_n)$ codes for which  
\begin{align}
\liminf_{n\to\infty}\frac{1}{n}\log M_n\ge R, \qquad \limsup_{n\to\infty}\epsilon_n \le \epsilon . \label{eqn:rate_def}
\end{align}
The {\em $\epsilon$-capacity} $C_{\epsilon}$ is the supremum of all $\epsilon$-achievable rates.  The {\em capacity} $C:=\lim_{\epsilon\to 0}C_\epsilon$. 
\end{definition}
The capacity of the general Gel'fand-Pinsker channel is stated in Section~\ref{sec:main_res}. This generalizes the result in~\eqref{eqn:gp_capacity}, which is the capacity for the  Gel'fand-Pinsker channel when the channel and state are stationary and memoryless. 

\subsection{The Gel'fand-Pinsker Problem With Coded State Information at the Decoder}
In fact the general information spectrum techniques allow us to solve a related problem which was first  considered by Heegard and El Gamal~\cite{heegard} and subsequently solved  by Steinberg~\cite{Ste08}.  The setup is shown in Fig.~\ref{fig:gp}.
\begin{definition} \label{def:coded_prob}
An {\em $(n,M_n,M_{\mathrm{d}, n}, \epsilon)$ code for the  Gel'fand-Pinsker problem} with channel $W^n:\calX^n\times\calS^n\to\calY^n$ and state distribution $P_{S^n} \in\calP(\calS^n)$ and with {\em coded state information} at the decoder consists of 
\begin{itemize}
\item A state encoder: $f_{\mathrm{d}, n}:\calS^n\to[1:M_{\mathrm{d}, n}]$
\item An encoder: $f_n: [1:M_n]\times\calS^n\to\calX^n$ (possibly randomized)
\item A decoder: $\varphi_n:\calY^n\times[1:M_{\mathrm{d}, n}]\to[1:M_n]$
\end{itemize}
such that the average error probability in decoding the message is no larger than $\epsilon$, i.e., 
\begin{equation}
 \frac{1}{M_n}\sum_{s^n\in\calS^n}P_{S^n}(s^n)\sum_{m=1}^{M_n}W^n( \calB_{m,s^n}^c  |f_n(m,s^n),s^n)\le\epsilon
\end{equation}
where $\calB_{m,s^n}:=\{y^n \in \calY^n:\varphi_n(y^n, f_{\mathrm{d}, n}(s^n))=m\}$.
\end{definition}
\begin{definition} \label{def:coded}
We say that the pair of nonnegative numbers $(R,\Rd) $ is {\em an achievable rate pair} if there exists a sequence of  $(n,M_n,M_{\mathrm{d}, n}, \epsilon_n)$ codes such that in addition to \eqref{eqn:rate_def},  the following holds
\begin{align}
 \limsup_{n\to\infty}\frac{1}{n}\log M_{\mathrm{d}, n}\le \Rd .  \label{eqn:defs_coded}
\end{align}
The set of all  achievable rate pairs is known as the  {\em capacity region} $\scC$. 
\end{definition}
Heegard and El Gamal~\cite{heegard} (achievability) and Steinberg~\cite{Ste08} (converse) showed for the discrete memoryless channel and discrete memoryless state   that the capacity region  $\scC$ is the set of rate pairs $(R,\Rd)$ such that 
\begin{align}
\Rd&\ge I(V;S)-I(V;Y) \label{eqn:wz_part}\\
R&\le I(U;Y|V)-I(U;S|V) \label{eqn:cgp_part}
\end{align}
for some Markov chain $(U,V) -(X,S)-Y$. Furthermore, $X$ can be taken to be a deterministic function of $(U,V,S)$, $|\calV|\le|\calX||\calS|+1$ and $|\calU|\le|\calX||\calS| (|\calX||\calS|+1)$. The constraint in~\eqref{eqn:wz_part} is obtained using Wyner-Ziv coding with ``source'' $S$ and ``side-information'' $Y$. The constraint in~\eqref{eqn:cgp_part} is analogous to the Gel'fand-Pinsker capacity where $V$ is common to both encoder and decoder. A weak converse was proven using repeated applications of the Csisz\'{a}r-sum-identity. We generalize Steinberg's region for the general source $\bS$ and general channel $\bW$ using information spectrum techniques in Section~\ref{sec:coded_dec}. 

\section{Information Spectrum Characterization of the General Gel'fand-Pinsker problem} \label{sec:info}
In this Section, we first present the main result concerning the capacity of the general Gel'fand-Pinsker problem (Definition~\ref{def:gp_prob})  in Section~\ref{sec:main_res}. These results are derived using the information spectrum method. We then derive the capacity for various special cases of the  Gel'fand-Pinsker problem  in Section~\ref{sec:corollaries} (two-sided common state information) and Section~\ref{sec:memoryless} (memoryless channels and state). We   consider mixed states and mixed channels in Section~\ref{sec:mixed}. The main ideas in the proof are discussed in Section~\ref{sec:idea}. Finally, in Section~\ref{sec:coded_dec}, we extend our result to the general  Gel'fand-Pinsker problem with coded state information at the decoder (Definition~\ref{def:coded_prob}).  

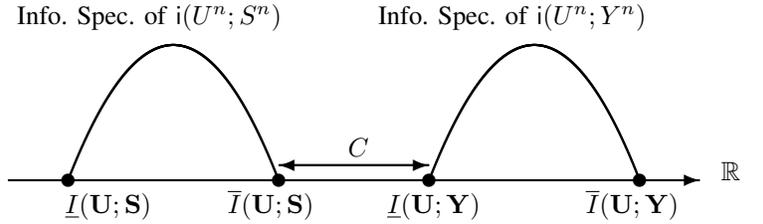
\begin{figure}
\centering
\setlength{\unitlength}{.4mm}
\begin{picture}(240, 65)

\put(20, 10){\circle*{4}}
\put(90, 10){\circle*{4}}
\put(140, 10){\circle*{4}}
\put(210, 10){\circle*{4}}
\thicklines
\put(0, 10){\vector(1, 0){230}}

\qbezier(20, 10)(55, 100)(90, 10)
\qbezier(140, 10)(175, 100)(210, 10)

\put(090, 15){\vector(1,0){50}}
\put(140, 15){\vector(-1,0){50}}

\put(110, 18){  $C$}
\put(234, 10){  $\bbR$}

%
%
%
\put(16, -1){  $\underI(\bU;\bS)$}
\put(70, -1){  $\overI(\bU;\bS)$}

\put(123, -1){  $\underI(\bU;\bY)$}
\put(189, -1){  $\overI(\bU;\bY)$}

\put(0, 62){ Info.\ Spec.\ of $\rvi(U^n;S^n)$}
\put(120,62){ Info.\ Spec.\  of $\rvi(U^n;Y^n)$}
%
%
%
%
  \end{picture}
  \caption{Illustration of Theorem~\ref{thm:gp_ach} where $\rvi(U^n;S^n):=n^{-1}\log [ {P_{U^n|S^n} (U^n|S^n)}/ {P_{U^n}(U^n)}]$ and similarly for $\rvi(U^n;Y^n)$. The capacity is the difference   between $\underI(\bU;\bY)$ and $\overI(\bU;\bS)$ evaluated at the optimal processes.  The stationary, memoryless case (Corollary~\ref{cor:stat}) corresponds to the situation in which $\underI(\bU;\bS)=\overI(\bU;\bS)=I(U;S)$ and $\underI(\bU;\bY)=\overI(\bU;\bY)=I(U;Y)$  so the information spectra are point masses at the mutual informations.  }
  \label{fig:2dist}
\end{figure}

\subsection{Main Result and Remarks} \label{sec:main_res}
We now state our main result followed by some simple remarks. The  proof can be found in Appendices~\ref{sec:prf:gp} and~\ref{sec:prf:lem}.
\begin{theorem}[General Gel'fand-Pinsker Capacity] \label{thm:gp_ach}
The  capacity of the general Gel'fand-Pinsker channel with general states $(\bW,\bS)$ (see Definition~\ref{def:gp}) is  
\begin{equation}
C = \sup_{\bU-(\bX,\bS)-\bY } \underI(\bU;\bY)-\overI(\bU;\bS) \label{eqn:gp_res}
\end{equation}
where the maximization is over all sequences of random variables $(\bU,\bX,\bS,\bY)=\{U^n , X^n, S^n, Y^n\}_{n=1}^{\infty}$ forming the requisite Markov chain,\footnote{For three processes $(\bX,\bY,\bZ)=\{ X^n, Y^n, Z^n\}_{n=1}^{\infty}$, we say that $\bX-\bY-\bZ$ {\em forms a Markov chain} if $X^n-Y^n-Z^n$ for all $n\in\bbN$.} having the state distribution coinciding with $\bS$ and having conditional distribution of $\bY$ given $(\bX,\bS)$ equal to  the general channel $\bW$. 
\end{theorem}
See Fig.~\ref{fig:2dist} for an illustration of Theorem~\ref{thm:gp_ach}.

\begin{remark} {\em 
The general formula in~\eqref{eqn:gp_res} is the dual of that in the Wyner-Ziv case~\cite{iwata02}. However, the proofs, and in particular, the  constructions of the codebooks, the notions of typicality and the application of Wyner's PBL, are subtly different from~\cite{miyake,iwata02, Kuz12}. We discuss these issues in Section~\ref{sec:idea}. Another problem which involves  difference of mutual information quantities is the wiretap channel~\cite[Chapter~22]{elgamal}. General formulas for the secrecy capacity using channel resolvability theory~\cite[Chapter~5]{Han10} were provided by Hayashi~\cite{Hayashi06} and Bloch and Laneman~\cite{bloch13}. They also involve the difference between spectral inf-mutual information rate (of the input and the legitimate receiver) and sup-mutual information rate (of the input and the eavesdropper).   
} \end{remark}

\begin{remark} {\em 
Unlike the usual Gel'fand-Pinsker formula for stationary and memoryless channels and states in~\eqref{eqn:gp_capacity}, we cannot conclude that the   conditional distribution we are optimizing over $P_{X^n,U^n|S^n}$ in~\eqref{eqn:gp_res} can be decomposed into the conditional distribution of $U^n$ given $S^n$ (i.e., $P_{U^n|S^n}$) and a deterministic function (i.e., $\bone\{x^n=g_n(u^n,s^n)\}$). }\end{remark}

\begin{remark}{\em 
If there is no state, i.e., $\calS = \emptyset$ in~\eqref{eqn:gp_res}, then we recover the general formula for channel capacity by Verd\'{u} and Han (VH)~\cite{VH94} 
\begin{equation}
C_{\mathrm{VH}} = \sup_{\bX } \underI(\bX;\bY ) .\label{eqn:vh}
\end{equation} 
The achievability follows by setting $\bU=\bX$. The converse follows by noting that $\underI(\bU;\bY)\le\underI(\bX;\bY)$ because $\bU-\bX-\bY$~\cite[Theorem 9]{VH94}. This is  the analogue of the data processing inequality for the spectral inf-mutual information rate. }
\end{remark}

\begin{remark}{\em  \label{rmk:cc}
The general  formula in~\eqref{eqn:gp_res} can be  slightly generalized to the Cover-Chiang (CC) setting~\cite{coverChiang} in which (i) the channel $W^n:\calX^n\times\calS_{\mathrm{e}}^n \times \calS_{\mathrm{d}}^n\to\calY^n$ depends on two state sequences $(S^n_{\mathrm{e}},S^n_{\mathrm{d}})\sim P_{S^n_{\mathrm{e}},S^n_{\mathrm{d}}}$ (in addition to $X^n$), (ii) partial channel state information $S^n_{\mathrm{e}}$ is available noncausally at the encoder and  (iii) partial  channel  state information $S^n_{\mathrm{d}}$ is  available at the decoder.  In this case, replacing $\bY$ with $(\bY, \bS_{\mathrm{d}})$ and $\bS$ with $\bS_{\mathrm{e}}$ in \eqref{eqn:gp_res}  yields
\begin{equation}
C_{\mathrm{CC}} = \sup_{\bU-(\bX,\bS_{\mathrm{e}},\bS_{\mathrm{d}})-(\bY, \bS_{\mathrm{d}}) } \underI(\bU;\bY,\bS_{\mathrm{d}})-\overI(\bU;\bS_{\mathrm{e}}) ,\label{eqn:gp_res_cc}
\end{equation}  
where the supremum is over all processes $(\bU,\bX,\bS_{\mathrm{e}},\bS_{\mathrm{d}},\bY)$ such that $(\bS_{\mathrm{e}},\bS_{\mathrm{d}})$ coincides with the state distributions  $\{P_{S^n_{\mathrm{e}},S^n_{\mathrm{d}}}\}_{n=1}^{\infty}$ and $\bY$ given $(\bX,\bS_{\mathrm{e}},\bS_{\mathrm{d}})$ coincides with the sequence of channels $\{W^n\}_{n=1}^{\infty}$. Hence the optimization  in~\eqref{eqn:gp_res_cc} is over the conditionals $\{ P_{X^n, U^n| S_{\mathrm{e}}^n}\}_{n=1}^{\infty}$.
 }
\end{remark}

\subsection{Two-sided Common State Information} \label{sec:corollaries}
Specializing~\eqref{eqn:gp_res_cc}  to the case where $\bS_{\mathrm{e}}=\bS_{\mathrm{d}} =\bS$, i.e., the {\em same} side information is available to both encoder and decoder (ED), does not appear to be straightforward without    further assumptions. Recall that in the usual scenario~\cite[Case~4 in~Corollary~1]{coverChiang}, we use the identification $U=X$ and chain rule for mutual information to assert that $I(X;Y,S)-I(X;S)=I(X;Y|S)$ evaluated at the optimal $P_{X|S}$ is the capacity. However, in information spectrum analysis, the chain rule does not hold for the $\pliminf$ operation. In fact, $\pliminf$ is superadditive~\cite{Han10}.     Nevertheless, under the assumption that a sequence of information densities  converges in probability, we can derive the capacity of the general channel with general common state available at both terminals using Theorem~\ref{thm:gp_ach}.
\begin{corollary}[General Channel Capacity with State at ED] \label{cor:both}
Consider the problem
\begin{equation}
 C_{\mathrm{ED}}=\sup_{\bX }\underI(\bX;\bY|\bS) , \label{eqn:both}
\end{equation}
where the supremum   is over all   $(\bX,\bS,\bY)$ such that  $\bS$ coincides with the given  state distributions $\{P_{S^n}\}_{n=1}^{\infty}$  and  $\bY$ given $(\bX,\bS)$ coincides with the  given channels $\{W^n\}_{n=1}^{\infty}$.  Assume that the maximizer  of~\eqref{eqn:both} exists and denote it by $\{P_{X^n|S^n}^*\}_{n=1}^{\infty}$. Let the distribution of $\bX^*$ given $\bS$ be $\{P_{X^n|S^n}^*\}_{n=1}^{\infty}$.  If
\begin{equation}
\underI(\bX^*;\bS)=\overI(\bX^*;\bS), \label{eqn:str_cov}
\end{equation}
then the capacity of the state-dependent channel with state   $\bS$ available at  both encoder and decoder is $C_{\mathrm{ED}}$ in~\eqref{eqn:both}. 
\end{corollary}
The proof is provided in Appendix~\ref{app:both}.  If the joint process $(\bX^*,\bS)$ satisfies~\eqref{eqn:str_cov} (and $\calX$ and $\calS$ are finite sets), it is called {\em information stable}~\cite{HV93}. In other words, the limit distribution of $n^{-1}\log   [ {P_{U^n|S^n} (U^n|S^n)}/ {P_{U^n}(U^n)}]$ (where $P_{U^n|S^n}=P_{X^n|S^n}^*$) in Fig.~\ref{fig:2dist} concentrates at a single point. We remark that a different achievability proof technique (that does not use Theorem~\ref{thm:gp_ach}) would allow us to dispense of the  information stability   assumption.  We can simply develop a conditional version of Feinstein's lemma  \cite[Lemma 3.4.1]{Han10} to prove the direct part of~\eqref{eqn:both}. However, we choose to start from Theorem~\ref{thm:gp_ach}, which is the most general capacity result for the Gel'fand-Pinsker problem.    Note that the converse of Corollary~\ref{cor:both} does not require~\eqref{eqn:str_cov}.  

\subsection{Memoryless Channels and Memoryless States} \label{sec:memoryless}
To see how we can use Theorem~\ref{thm:gp_ach} in concretely,  we specialize it to the memoryless (but not necessarily  stationary) setting and we provide some interesting examples. In the memoryless setting, the sequence of channels  $\bW=\{W^n\}_{n=1}^{\infty}$ and  the sequence of state distributions $P_{\bS}=\{P_{S^n}\}_{n=1}^{\infty}$ are such that for every $(x^n, y^n, s^n)\in\calX^n\times\calY^n\times\calS^n$, we have $W^n(y^n|x^n,s^n)=\prod_{i=1}^n W_i(y_i|x_i, s_i)$, and  $P_{S^n}(s^n)=\prod_{i=1}^n P_{S_i}(s_i)$ for some $\{W_i:\calX\times\calS\to\calY\}_{i=1}^{\infty}$ and some  $\{P_{S_i} \in\calP(\calS)\}_{i=1}^{\infty}$. 
\begin{corollary}[Memoryless Gel'fand-Pinsker Channel Capacity] \label{cor:mem}
Assume that $\calX,\calY$ and $\calS$ are finite sets and the Gel'fand-Pinsker channel is memoryless and characterized by $\{W_i\}_{i=1}^{\infty}$  and $\{P_{S_i}\}_{i=1}^{\infty}$. Define $\phi( P_{X,U|S}; W, P_S):=I(U;Y)-I(U;S)$. Let the maximizers  to the optimization problems  indexed by $i\in\bbN$ 
\begin{equation}
C(W_i,P_{S_i})= \max_{\substack{P_{X,U|S} \\|\calU|\le|\calX| |\calS|+1}}\phi(  P_{X,U|S}; W_i, P_{S_i})\label{eqn:ith_opt}
\end{equation}
be denoted as $P_{X_i,U_i|S_i}^*:\calS\to\calX\times\calU$. Let $P_{S_i, X_i, U_i, Y_i}^*=P_{S_i}\circ P_{X_i,U_i|S_i}^*\circ W^n_i \in \calP(\calS\times\calX\times\calU\times\calY)$ be the joint distribution induced by $P_{X_i,U_i|S_i}^*$. Assume that either of the two limits
\begin{equation}
  \lim_{n\to\infty}\frac{1}{n}\sum_{i=1}^n I(P_{S_i},P_{U_i|S_i}^* ), \quad   \lim_{n\to\infty}\frac{1}{n}\sum_{i=1}^n I(P_{U_i}^*,P_{Y_i|U_i}^* )\label{eqn:lims_exist}
\end{equation}
exist.  Then  the capacity of the memoryless Gel'fand-Pinsker channel is 
\begin{equation}
C_{\mathrm{M'less}}=\liminf_{n\to\infty}\frac{1}{n}\sum_{i=1}^n C(W_i, P_{S_i}). \label{eqn:mless}
\end{equation}
\end{corollary}
The proof of Corollary~\ref{cor:mem} is detailed in Appendix~\ref{app:converse}.  The  Ces\`aro  summability assumption in~\eqref{eqn:lims_exist}      is  only required for achievability. We illustrate the   assumption in \eqref{eqn:lims_exist} with two examples in the sequel. The proof of the direct part of Corollary~\ref{cor:mem} follows by taking the optimization in the general result~\eqref{eqn:gp_res} to be over memoryless  conditional distributions. The converse follows by  repeated  applications of the Csisz\'ar-sum-identity~\cite[Chapter 2]{elgamal}.  If in addition to being memoryless, the channels and states are stationary (i.e., each $W_i$ and each $P_{S_i}$   is equal to $W$ and $P_S$ respectively), the both limits in~\eqref{eqn:lims_exist}  exist since $P_{X_i,U_i|S_i}^*$ is the same for each $i\in\bbN$. 
\begin{corollary}[Stationary, Memoryless Gel'fand-Pinsker Channel Capacity]  \label{cor:stat}
Assume that $\calS$ is a finite set. In the stationary, memoryless case, the capacity  of the   Gel'fand-Pinsker channels given by $C(W,P_S)$  in~\eqref{eqn:gp_capacity}.
\end{corollary}
We omit the proof because it is a straightforward consequence of  Corollary~\ref{cor:mem}. Close examination of the proof of Corollary~\ref{cor:mem} shows that only the converse of Corollary~\ref{cor:stat} requires the assumption that $|\calS|<\infty$. The achievability    of  Corollary~\ref{cor:stat}  follows easily from  Khintchine's law of large numbers~\cite[Lemma 1.3.2]{Han10} (for abstract alphabets).

To gain a better understanding of the assumption in~\eqref{eqn:lims_exist} in Corollary~\ref{cor:mem}, we now present a couple of (pathological) examples which are inspired by~\cite[Remark 3.2.3]{Han10}.

\begin{example} \label{eg:no_converge}
{\em  Let $\calJ:=\{i\in\bbN: 2^{2k-1}\le i < 2^{2k}, k \in \bbN\}= [2:3]\cup[8:15]\cup[32:63]\cup\ldots$. Consider a discrete,  nonstationary, memoryless channel $\bW$  satisfying
 \begin{align}
W_i=\left\{ \begin{array}{cc}
\tilW_{\mathrm{a}} & i\in\calJ \\
\tilW_{\mathrm{b}} & i\in\calJ^c 
\end{array}\right.  ,
\end{align}
where $\tilW_{\mathrm{a}},\tilW_{\mathrm{b}} :\calX\times\calS\to\calY$ are two distinct  channels. Let $P_{S^n}=  Q^n$ be the $n$-fold extension of some  $Q\in\calP(\calS)$. Let $V_{m}^*:\calS\to\calU$ be the $\calU$-marginal of the maximizer of \eqref{eqn:ith_opt} when the channel is $\tilW_{ {m}},m\in\{\rma,\rmb\}$.    In general, $I(Q,V_{\rma}^*)\ne I(Q,  V_{\rmb}^*)$. Because  $\liminf_{n\to\infty} \frac{1}{n}|\calJ\cap [1:n]|=\frac{1}{3}\ne\limsup_{n\to\infty} \frac{1}{n}|\calJ\cap [1:n]|=\frac{2}{3}$, the first limit in \eqref{eqn:lims_exist} does not exist. Similarly, the second  limit does not exist in general  and Corollary~\ref{cor:mem} cannot be applied. 
}
\end{example}
\begin{example} \label{eg:nonerg}
{\em Let $\calJ$ be as in Example~\ref{eg:no_converge} and let the set of even and odd positive integers be $\calE$ and $\calO$ respectively.  Let  $\calS,\calX,\calY=\{0,1\}$. Consider a binary, nonstationary, memoryless channel $\bW$ satisfying 
\begin{align}
W_i=\left\{ \begin{array}{cc}
\tilW_{\mathrm{a}} & i\in\calO \cap \calJ\\
\tilW_{\mathrm{b}} & i\in\calO\cap \calJ^c \\
\tilW_{\mathrm{c}} & i\in\calE 
\end{array}\right.  ,  \label{eqn:ch_eg}
\end{align}
where $\tilW_{\mathrm{a}},\tilW_{\mathrm{b}},\tilW_{\mathrm{c}}:\calX\times\calS\to\calY$. Also consider a   binary, nonstationary, memoryless state $\bS$ satisfying
\begin{align}
P_{S_i}=\left\{ \begin{array}{cc}
Q_{\mathrm{a}} & i\in\calO\\
Q_{\mathrm{b}} & i\in\calE
\end{array}\right. , \label{eqn:st_eg}
\end{align}
where $Q_{\mathrm{a}}, Q_{\mathrm{b}} \in \calP(\{0,1\})$.  In addition, assume that $\tilW_m(\fndot|\fndot,s)$   for $(m,s) \in \{\mathrm{a},\mathrm{b}\}\times\{0,1\}$  are binary symmetric channels with  arbitrary crossover probabilities $q_{ms}\in (0,1)$.  Let $V_{m,l}^*:\calS\to\calU$ be the $\calU$-marginal of the  maximizer in~\eqref{eqn:ith_opt} when the channel is $\tilW_{ {m}},m\in\{\rma,\rmb\}$ and the state distribution is $Q_l,l\in \{\rma,\rmb\}$. For $m \in \{\mathrm{a},\mathrm{b}\}$ (odd blocklengths), due to the symmetry of the channels the optimal $V_{m,\rma}^*(u|s)$ is Bernoulli$(\frac{1}{2})$   and independent of $s$~\cite[Problem~7.12(c)]{elgamal}.  Thus, for all odd blocklengths, the mutual informations in the first limit in~\eqref{eqn:lims_exist} are  equal to zero. Clearly, the  first  limit in~\eqref{eqn:lims_exist} exists, equalling $\frac{1}{2}I(Q_{\rmb}, V_{\rmc,\rmb}^*)$ (contributed by the even blocklengths). Therefore, Corollary~\ref{cor:mem} applies and we can show that  the Gel'fand-Pinsker capacity is  
\begin{equation}
C=\frac{1}{2} \left[ G(Q_{\mathrm{a}})+C(\tilW_{\mathrm{c}},Q_{\mathrm{b}}) \right] \label{eqn:halfhalf}
\end{equation}
where $C(W,P_S)$ in~\eqref{eqn:ith_opt} is given explicitly in~\cite[Problem~7.12(c)]{elgamal}   and $G:\calP(\{0,1\})\to\bbR$ is defined as 
\begin{align}
G(Q)&:=\frac{2}{3}\min\left\{  C(\tilW_{\mathrm{a}}, Q), C(\tilW_{\mathrm{b}}, Q) \right\}  \nn\\
&\quad +\frac{1}{3}\max\left\{ C(\tilW_{\mathrm{a}}, Q), C(\tilW_{\mathrm{b}}, Q)\right\} . \label{eqn:CQ}
\end{align}
See Appendix~\ref{app:example} for the derivation of~\eqref{eqn:halfhalf}. The   expression in \eqref{eqn:halfhalf} implies that the capacity consists of two parts: $C(\tilW_{\mathrm{c}},Q_{\mathrm{b}})$ represents the performance of the system  $(\tilW_{\mathrm{c}},Q_{\mathrm{b}})$ at even blocklengths, while $G(Q_{\mathrm{a}})$ represents the non-ergodic behavior of the channel  at odd blocklengths with state distribution $Q_{\rma}$; cf.~\cite[Remark 3.2.3]{Han10}. In the special case that $C(\tilW_{\mathrm{a}}, Q_{\mathrm{a}}) = C(\tilW_{\mathrm{b}}, Q_{\mathrm{a}})$ (e.g., $\tilW_{\mathrm{a}}=\tilW_{\mathrm{b}}$), then the  capacity is   the average of $G(Q_{\mathrm{a}})=C(\tilW_{\mathrm{a}},Q_{\mathrm{a}})$  and  $C(\tilW_{\mathrm{c}},Q_{\rmb })$. 
}
\end{example}

\subsection{Mixed Channels and Mixed States} \label{sec:mixed}
Now we use Theorem~\ref{thm:gp_ach} to compute the capacity of the Gel'fand-Pinsker channel when the channel and state sequence are {\em mixed}. More precisely, we assume that
\begin{align}
W^n(y^n|x^n, s^n)&=\sum_{k=1}^{\infty}\alpha_k W_k^n(y^n|x^n, s^n), \label{eqn:mix_W} \\
 P_{S^n}(s^n)&=\sum_{l=1}^{\infty}\beta_l P_{S_l^n}(s^n).\label{eqn:mix_S}
\end{align}
Note that we require $\sum_{k=1}^{\infty}\alpha_k=\sum_{l=1}^{\infty}\beta_l=1$. In fact, let   $\calK:=\{k\in\bbN:\alpha_k>0\}$ and $\calL:=\{l\in\bbN:\beta_l>0\}$. Note that if $S_l^n$ is a stationary and memoryless source, $S^n$  the composite source given by~\eqref{eqn:mix_S}, is  a canonical example of a {\em non-ergodic} and stationary  source. By~\eqref{eqn:mix_W}, the channel $W^n$ can be regarded as an average channel given by the convex combination of $|\calK|$   constituent channels. It is stationary but {\em non-ergodic} and {\em non-memoryless}. Given $P_{X^n,U^n|S^n}$,   define the following random variables which are indexed by $k$ and $l$:
\begin{equation}
(S^n_{l}, X^n_{l}, U^n_{l}, Y^n_{kl})\sim P_{S_l^n} \circ P_{X^n,U^n|S^n}\circ W_k^n.\label{eqn:product_mixed}
\end{equation}

\begin{corollary}[Mixed  Channels and Mixed States] \label{cor:mixed1}
The  capacity of the general mixed Gel'fand-Pinsker channel with general  mixed state as in~\eqref{eqn:mix_W}--\eqref{eqn:product_mixed} is
\begin{equation}
C = \sup_{\bU-(\bX,\bS)-\bY }  \left\{\inf_{(k,l)\in\calK\times\calL}\underI(\bU_{l};\bY_{kl})-\sup_{l\in \calL}\overI(\bU_{l};\bS_{l}) \right\} \label{eqn:gp_res_mixed}
\end{equation}
where the maximization is over all sequences of random variables $(\bU,\bX,\bS,\bY)=\{U^n , X^n, S^n, Y^n\}_{n=1}^{\infty}$ with state distribution coinciding with $\bS$ in~\eqref{eqn:mix_S} and having conditional distribution of $\bY$ given $(\bX,\bS)$ equal to  the general channel $\bW$ in~\eqref{eqn:mix_W}.   Furthermore, if each general state sequence $S_l^n$ and each general channel $W_k^n$ is stationary and memoryless, the capacity is lower bounded as 
\begin{equation}
C \ge \max_{U-(X,S)-Y}  \left\{\inf_{(k,l)\in\calK\times\calL}I(U_{l};Y_{kl})-\sup_{l\in \calL}I(U_{l};S_{l}) \right\} \label{eqn:gp_res_mixed_st}
\end{equation}
where  $(S_l,X_l,U_l,Y_{kl})\sim P_{S_l}\circ P_{X,U|S}\circ W_k$ and the maximization is over all joint distributions $P_{U,X,S,Y}$ satisfying $P_{U,X,S,Y}=\sum_{k,l}\alpha_k\beta_l P_{S_l}P_{X,U|S}W_k$ for some $P_{X,U|S}$. 
\end{corollary}
Corollary~\ref{cor:mixed1} is proved in Appendix~\ref{sec:prf:cor1} and it basically applies~\cite[Lemma 3.3.2]{Han10} to the mixture with components in~\eqref{eqn:product_mixed}. Different from existing analyses for mixed channels and sources~\cite{Han10,iwata02}, here there are two {\em independent}  mixtures---that of the channel {\em and} the state. Hence, we need to minimize  over two indices for the first term in~\eqref{eqn:gp_res_mixed}.    If  instead of the countable number of terms in the sums in~\eqref{eqn:mix_W} and~\eqref{eqn:mix_S}, the number of mixture components (of either the source or channel) is  uncountable, Corollary~\ref{cor:mixed1} no longer applies   and a corresponding  result has to involve the assumptions that the alphabets are finite and the constituent  channels are memoryless. See \cite[Theorem 3.3.6]{Han10}.

 The corollary says that the Gel'fand-Pinsker capacity  is governed by two elements: (i) the ``worst-case'' virtual channel (from $\calU^n$ to $\calY^n$), i.e., the one with the smallest packing rate $\underI(\bU_{l};\bY_{kl})$  and (ii) the ``worst-case'' state distribution, i.e., the one that results in the largest covering rate $ \overI(\bU_l;\bS_l)$.   Unfortunately,    obtaining a converse result for the stationary, memoryless case from~\eqref{eqn:gp_res_mixed} does not appear to be straightforward. The same issue was also encountered for the mixed wiretap channel~\cite{bloch13}.

\subsection{Proof Idea of Theorem~\ref{thm:gp_ach}} \label{sec:idea}
\subsubsection{Direct part}\label{sec:idea_d}
The high-level idea in the achievability proof  is similar to  the usual Gel'fand-Pinsker coding scheme~\cite{GP80} which involves a covering step to reduce the uncertainty due to the random state sequence and a packing step to decode the transmitted codeword.  However,   to use the information spectrum method on the general channel and general  state, the definitions of ``typicality''   have to be restated in terms of information densities. See the definitions in Appendix~\ref{sec:prf:gp}. The main burden for the proof  is to show that the probability that the transmitted codeword $U^n$ is not ``typical'' with the channel output $Y^n$ vanishes. In regular Gel'fand-Pinsker coding, one appeals to  the conditional typicality lemma \cite[Lemma 2]{GP80}  \cite[Chapter 2]{elgamal}     (which holds for ``strongly typical sets'') to assert that this error probability is small. But the ``typical sets'' used in information spectrum analysis do not allow us to apply the conditional typicality lemma  in a straightforward manner. For example, our decoder is a threshold test involving the information density statistic $n^{-1}\log (P_{Y^n|U^n}/P_{Y^n})$. It is not clear in the event that there is no covering error that the transmitted $U^n$ codeword   passes the threshold test (i.e., $n^{-1}\log (P_{Y^n|U^n}/P_{Y^n})$ exceeds a certain threshold)  with high probability.    

To get around this problem, we   modify  Wyner's PBL\footnote{ \label{fn:pbl} One    version of the piggyback coding lemma (PBL), given in~\cite[Lemma A.1]{rag13}, can be stated as follows: If $U-V-W$ are   random variables forming Markov chain, $(U^n,V^n,W^n)\sim\prod_{i=1}^n P_{U,V}(u_i,v_i)P_{W|V}( w_i|v_i)$, and $\psi_n:\calU^n \times\calW^n\to [0,1]$ is a function satisfying $\bbE\psi_n(U^n,W^n)\to 0$, then for any given $\veps>0$, for all sufficiently large $n$ there exists a mapping $g_n:\calV^n\to\calW^n$ such that (i) $\frac{1}{n}\log\|g_n\|\le I(V;W)+\veps$ and (ii) $\bbE\psi_n(U^n, g_n(V^n))<\veps$. The function $\psi_n(u^n,w^n)$ is usually taken to be the indicator that $(u^n,w^n)$ are jointly typical.}~\cite[Lemma 4.3]{Wyner75} \cite[Lemma A.1]{rag13}  accordingly. Wyner  essentially derived an analog of the Markov lemma~\cite[Chapter~12]{elgamal}    without strong typicality by  introducing a new ``typical set''  defined in terms of conditional probabilities. This new definition is particularly useful  for  problems that involve   covering and packing as well as having  some Markov structure.    Our analysis is somewhat similar to the analyses of the general Wyner-Ziv problem  in~\cite{iwata02} and  the  WAK problem in~\cite{miyake, Kuz12}. This is unsurprising given that the Wyner-Ziv and Gel'fand-Pinsker problems are duals~\cite{coverChiang}. However, unlike in~\cite{iwata02}, we   construct   random subcodebooks and use them in subsequent steps rather than to assert the existence of a single codebook via random selection and subsequently regard it as being deterministic.    This is because unlike Wyner-Ziv, we need to construct exponentially many subcodebooks each   of size $\approx\exp(n \overI(\bU;\bS))$ and indexing a message in $[1:M_n]$.    We also require each of these subcodebooks to be {\em different} and {\em identifiable} based on the channel output.   Also, our analogue of   Wyner's ``typical set''  is different  from previous works.   

We also point out that Yu {\em et al.}~\cite{WeiYu} considered the Gaussian  Gel'fand-Pinsker problem for non-stationary  and non-ergodic channel and state. However, the notion of typicality used  is ``weak typicality'', which means that the sample entropy is close to the entropy rate. This notion does not   generalize well for obtaining the capacity expression in~\eqref{eqn:gp_res}, which involves   limits in probability of information densities. Furthermore, Gaussianity is a crucial hypothesis in the proof of the asymptotic equipartition property in~\cite{WeiYu}. 


\subsubsection{Converse part}
For the converse, we use  the Verd\'{u}-Han   converse~\cite[Lemma 3.2.2]{Han10} and the fact that the message is independent of the state sequence.  Essentially, we emulate the steps for the converse of the general wiretap channel presented by Bloch and Laneman~\cite[Lemma~3]{bloch13}. 

\subsection{Coded State Information at Decoder} \label{sec:coded_dec}
We now state  the capacity region of the coded state information problem (Definition~\ref{def:coded_prob}). 
\begin{theorem}[Coded State Information at Decoder]  \label{thm:coded}
The capacity region of the  Gel'fand-Pinsker problem with coded state information at the decoder $\scC$  (see Definition~\ref{def:coded}) is given by the set of pairs $(R,\Rd)$ satisfying
\begin{align}
R&\le \underI(\bU;\bY|\bV)-\overI(\bU;\bS|\bV) \label{eqn:cgp}\\
\Rd&\ge \overI(\bV;\bS)-\underI(\bV;\bY) \label{eqn:wz}
\end{align}
for $(\bU,\bV,\bX,\bS,\bY)=\{U^n ,V^n, X^n, S^n, Y^n\}_{n=1}^{\infty}$ satisfying $(\bU,\bV)-(\bX,\bS)-\bY$, having the state distribution coinciding with $\bS$ and having conditional distribution of $\bY$ given $(\bX,\bS)$ equal to  the general channel $\bW$.  
\end{theorem}
A proof sketch is provided in Appendix~\ref{sec:prf:coded}. For the direct part, we   combine   Wyner-Ziv     and Gel'fand-Pinsker coding to obtain the two constraints in Theorem~\ref{thm:coded}. To prove the converse, we use exploit the independence of the message and the state, the Verd\'{u}-Han lemma~\cite[Lemma 3.2.2]{Han10} and the proof technique for the converse of the general   rate-distortion problem~\cite[Section 5.4]{Han10}.  Because the proof of Theorem~\ref{thm:coded} is very similar to Theorem~\ref{thm:gp_ach}, we only provide a sketch. We   note that similar ideas can be easily employed to find the  general capacity region for the problem of coded state information at the {\em encoder} (and full state information at the decoder) \cite{Rosen05}.  In analogy to Corollary~\ref{cor:stat}, we can use Theorem~\ref{thm:coded} to recover Steinberg's result~\cite{Ste08} for the stationary, memoryless  case. See Appendix~\ref{app:coded_sm}  for the proof.
\begin{corollary}[Coded State Information at Decoder for Stationary, Memoryless Channels and States]  \label{cor:coded}
Assume that $\calS$ is a finite set. The capacity of the  Gel'fand-Pinsker channel with coded state information at the decoder in the stationary, memoryless case is given   in~\eqref{eqn:wz_part} and~\eqref{eqn:cgp_part}.
\end{corollary}

\section{Conclusion} \label{sec:concl}
In this work, we derived the capacity of the general Gel'fand-Pinsker channel with general state distribution using the information spectrum method. We also extended the analysis to the case where coded state information is available at the decoder. 


\appendix
 
\subsection{Proof of Theorem~\ref{thm:gp_ach}} \label{sec:prf:gp}

{\em Basic definitions}: Fix $\gamma_1,\gamma_2>0$  and some conditional distribution $P_{X^n, U^n|S^n}$. Define the sets
\begin{align}
\calT_1 &:=\bigg\{ (u^n,y^n)\in\calU^n\times\calY^n: \frac{1}{n}\log \frac{P_{Y^n|U^n}(y^n|u^n)}{P_{Y^n}(y^n)} \nn\\*
 &\qquad \qquad \ge\underI(\bU;\bY)- \gamma_1 \bigg\} \label{eqn:calT1} \\*
\calT_2 &:=\bigg\{ (u^n,s^n)\in\calU^n\times\calS^n :\frac{1}{n}\log \frac{P_{U^n|S^n}(u^n|s^n)}{P_{U^n}(u^n)} \nn\\*
&\qquad \qquad \le\overI(\bU;\bS)+ \gamma_2 \bigg\}  \label{eqn:calT2} ,
\end{align}
where the random variables $(S^n,X^n,U^n,Y^n) \sim P_{S^n}\circ P_{X^n,U^n|S^n}\circ W^n$.  We define the probabilities
\begin{align}
\pi_1&:=\bbP( (U^n,Y^n)\notin\calT_1 ) \label{eqn:defpi1}\\
\pi_2&:=\bbP( (U^n,S^n)\notin\calT_2 ),\label{eqn:defpi2}
\end{align}
where   $(U^n,Y^n)\sim P_{U^n, Y^n}$ and   $(U^n,S^n)\sim P_{U^n, S^n}$ and  where these joint distributions are computed with respect to $P_{S^n}\circ P_{X^n,U^n|S^n}\circ W^n$.  Note that $\pi_1$ and $\pi_2$ are information spectrum~\cite{Han10} quantities.

\begin{proof}  We begin with  achievability.  We show that the rate 
\begin{equation}
R := \underI(\bU;\bY)-2\gamma_1-(\overI(\bU;\bS)+2\gamma_2) \label{eqn:rateR} 
\end{equation}
is achievable.  The next lemma provides an upper bound on the error probability in terms of the  above quantities. 

\begin{lemma}[Nonasymptotic upper bound on error probability for Gel'fand-Pinsker]\label{lem:bound_err}
Fix a sequence of conditional distributions $\{P_{X^n,U^n|S^n}\}_{n=1}^{\infty}$. This specifies $\underI(\bU;\bY)$ and $\overI(\bU;\bS)$. For every positive integer $n$,  there exists an $(n, \exp(nR), \rho_n)$ code for the general Gel'fand-Pinsker channel where $R$ is defined in~\eqref{eqn:rateR} and  
\begin{equation}
\rho_n := 2\pi_1^{1/2}+\pi_2+\exp \left( -\exp( n \gamma_2 ) \right)+\exp \left(-n  \gamma_1 \right)  . \label{eqn:bound_err}
\end{equation}
\end{lemma}
The proof of Lemma~\ref{lem:bound_err} is provided in Appendix~\ref{sec:prf:lem}. We note that there have been stronger nonasymptotic bounds for the Gel'fand-Pinsker problem developed recently~\cite{verdu12, WKT13, YAG13b} but Lemma~\ref{lem:bound_err} suffices for our purposes because the main theorem is an asymptotic one.  We prove Lemma~\ref{lem:bound_err} in full in Section~\ref{sec:prf:lem}. This argument  is classical and follows the original idea of Wyner's piggyback coding lemma~\cite[Lemma~4.3]{Wyner75} with a few modifications as discussed in Section~\ref{sec:idea_d}.  

Now, for any fixed $\gamma_1,\gamma_2>0$, the last two terms  in~\eqref{eqn:bound_err} tend to zero. By the definition of spectral inf-mutual information rate, 
\begin{equation}
\pi_1=\bbP\left(\frac{1}{n}\log \frac{P_{Y^n|U^n}(Y^n|U^n)}{P_{Y^n}(Y^n)}<\underI(\bU;\bY)- \gamma_1\right) 
\end{equation}
goes to zero. By the definition of the  spectral  sup-mutual information rate,
\begin{equation}
\pi_2=\bbP\left(\frac{1}{n}\log \frac{P_{U^n|S^n}(U^n|S^n)}{P_{U^n}(U^n)}>\overI(\bU;\bS)+ \gamma_2 \right)
\end{equation}
also goes to zero. Hence, in view of~\eqref{eqn:bound_err}, the error probability vanishes with increasing blocklength.   This proves that the rate $R$ in~\eqref{eqn:rateR} is achievable. Taking $\gamma_1,\gamma_2\to 0$ and maximizing over all chains $\bU-(\bX,\bS)-\bY$ proves  the direct part of Theorem~\ref{thm:gp_ach}.

For the converse, we follow the strategy for proving the converse for the general wiretap channel  as done by Bloch and Laneman~\cite[Lemma~3]{bloch13}. Consider a sequence of $(n,M_n, \epsilon_n)$   codes (Definition~\ref{def:gp_prob}) achieving a rate $R_n=\frac{1}{n}\log M_n$.  Let $U^n \in\calU^n$ denote an arbitrary random variable representing the uniform choice of a message in $[1:M_n]$.  Because the message is independent of the state (cf.~discussion after Definition~\ref{def:gp_prob}), this induces the joint distribution $P_{S^n} \circ P_{U^n} \circ  P_{X^n|U^n,S^n}\circ  W^n$ where $P_{X^n|U^n,S^n}$ models possible stochastic encoding. Clearly by the independence, 
\begin{equation}
\overI(\bU;\bS)=0. \label{eqn:indepUS}
\end{equation}
Let the set of processes $(\bS,\bU,\bX,\bY)$ in which each collection of random variables $(S^n, U^n, X^n, Y^n)$ is distributed as $P_{S^n} \circ P_{U^n} \circ  P_{X^n|U^n,S^n}\circ  W^n$    (resp.\ $P_{S^n} \circ P_{U^n|S^n} \circ  P_{X^n|U^n,S^n}\circ  W^n$) be $\calI_{\bW,\bS}$ for ``independent'' (resp.\ $\calD_{\bW,\bS}$ for ``dependent'').   Fix $\gamma>0$. The Verd\'{u}-Han converse theorem~\cite[Lemma 3.2.2]{Han10} states that for any $(n,M_n,\epsilon_n)$   code for the general virtual channel $P_{Y^n|U^n}(y^n|u^n):=\sum_{x^n,s^n} W^n(y^n|x^n , s^n) P_{X^n|U^n,S^n    }(x^n|u^n,s^n) P_{S^n}(s^n)$, 
\begin{align}
\epsilon_n&\ge\bbP\left(\frac{1}{n}\log\frac{P_{Y^n|U^n}(Y^n|U^n)}{P_{Y^n}(Y^n)}\le\frac{1}{n}\log M_n-\gamma\right) \nn\\*
 &\qquad -\exp(-n\gamma),
\end{align}
where $U^n$ is uniform over the message set $[1:M_n]$. Suppose now that  $M_n= \lceil\exp [ n (\underI(\bU;\bY)+2\gamma)] \rceil$. Then,
\begin{align}
\epsilon_n &\ge\bbP\left(\frac{1}{n}\log\frac{P_{Y^n|U^n}(Y^n|U^n)}{P_{Y^n}(Y^n)}\le\underI(\bU;\bY)+ \gamma\right) \nn\\*
  &\qquad -\exp(-n\gamma). \label{eqn:vh1}
\end{align}
 By the definition of the spectral inf-mutual information rate, the first term on the right hand side of~\eqref{eqn:vh1} converges to $1$. Since $\exp(-n\gamma)\to 0$, $\epsilon_n\to 1$ if $M_n=\lceil\exp [ n (\underI(\bU;\bY)+2\gamma)]\rceil$. This means that a necessary condition for the code to have vanishing error probability is  for  $R=\lim_{n\to\infty}R_n$ to satisfy
\begin{align}
R&\le\underI(\bU;\bY)+2\gamma \label{eqn:use_vh}\\
&= \underI(\bU;\bY)-\overI(\bU;\bS)+2\gamma \label{eqn:useindepUS}\\
&\le \sup_{(\bS,\bU,\bX,\bY)\in \calI_{\bW,\bS}  } \left\{ \underI(\bU;\bY)-\overI(\bU;\bS) \right\}+2\gamma, \label{eqn:mess_arb} \\
&\le \sup_{ (\bS,\bU,\bX,\bY)\in \calD _{\bW,\bS} }\left\{ \underI(\bU;\bY)-\overI(\bU;\bS) \right\} +2\gamma , \label{eqn:large_sup}
\end{align}
where~\eqref{eqn:useindepUS} follows from~\eqref{eqn:indepUS} and~\eqref{eqn:large_sup} follows because $\calI_{\bW,\bS}\subset\calD_{\bW,\bS}$ because the set of dependent processes  includes the independent processes as a special case. Since $\gamma>0$  is arbitrary, we have proven the upper bound of~\eqref{eqn:gp_res} and this completes the proof of Theorem~\ref{thm:gp_ach}. \end{proof}

\subsection{ Proof of Lemma~\ref{lem:bound_err}}\label{sec:prf:lem}

\begin{proof}
Refer to the definitions of the sets $\calT_1$ and $\calT_2$ and the probabilities $\pi_1$ and $\pi_2$ in Appendix~\ref{sec:prf:gp} in which the random variables $(S^n, X^n, U^n, Y^n)\sim P_{S^n}\circ P_{X^n,U^n|S^n}\circ W^n$. Define the mapping $\eta:\calU^n\times\calS^n\to \bbR_+$  as
\begin{align}
\eta(u^n,s^n):=\sum_{x^n}& \sum_{y^n: (u^n, y^n)\notin\calT_1} W^n(y^n|x^n,s^n)  \nn\\*
 & \times P_{X^n|U^n,S^n}(x^n|u^n,s^n) \label{eqn:eta_def}
\end{align}
where $ P_{X^n|U^n,S^n}$ is  the conditional induced by  $P_{X^n,U^n|S^n}$. Analogous to~\cite[Lemma 4.3]{Wyner75}, define the set 
\begin{equation}
\calA:= \left\{ (u^n,s^n)\in\calU^n\times\calS^n:\eta(u^n,s^n)\le \pi_1^{1/2}  \right\} .\label{eqn:defcalA}
\end{equation} 
Note the differences between the definition of $\eta$  vis-\`a-vis that in~\cite[Lemma 4.3]{Wyner75}. In particular, the summand in the definition of $\eta$ in~\eqref{eqn:eta_def} depends on $P_{X^n|U^n,S^n}$ and the channel $W^n$.   Now, for brevity, define the ``inflated rate''
\begin{equation}
\tilR:=\underI(\bU;\bY)-2\gamma_1\label{eqn:iflate}
\end{equation}
so the rate of each subcodebook is $\tilR-R = \overI(\bU;\bS)+2\gamma_2$. 

{\em Random code generation}: Randomly and independently generate $\lceil\exp(n\tilR) \rceil$ codewords $\{u^n(l): l\in [1:\exp(n\tilR)]\}$ each drawn  from $P_{U^n}$. Denote  the set of random codewords and a specific realization as  $\calC$ and $\frakc$ respectively. Deterministically partition the  codewords in $\calC$ into $\lceil\exp(nR)\rceil$ subcodebooks $\calC(m)=\{u^n(l):l\in [(m-1)\exp( n (\tilR-R))+1:m\exp( n (\tilR-R))]\}$ where $ m\in [1:\exp(nR)]$. Note that each subcodebook contains $\lceil\exp(n (\overI(\bU;\bS)+2\gamma_2)) \rceil$ codewords. Here is where our coding scheme differs from the general Wyner-Ziv problem~\cite{iwata02} and the general WAK problem~\cite{miyake, Kuz12}. We randomly generate exponentially many subcodebooks instead of asserting the existence of {\em one} by random selection via Wyner's PBL~\cite[Lemma 4.3]{Wyner75}. By retaining the randomness in the $U^n$ codewords, it is easier to bound the probability of the decoding error $\calE_3$  defined in~\eqref{eqn:calE_3}. 

{\em Encoding}: The encoder, given $m\in [1:\exp(nR)]$ and the state sequence $s^n\in\calS^n$ (noncausally), finds the sequence $u^n(\hatl)\in\calC(m)$ with the smallest index $\hatl$ satisfying
\begin{equation}
 (u^n(\hatl),s^n)\in\calA. \label{eqn:covering}
\end{equation}
If no such $\hatl$ exists, set $\hatl=1$. Randomly generate a sequence $x^n\sim P_{X^n|U^n,S^n}(\fndot|u^n(\hatl),s^n)$ and transmit it as the channel input in addition to $s^n$. Note that the rate of the code is given by $R$ in~\eqref{eqn:rateR} since there are $\lceil\exp(nR)\rceil$ subcodebooks, each representing one message.

{\em Decoding}: Given $y^n\in\calY^n$ decoder declares that $\hatm\in [1:\exp(nR)]$ is the message sent if it is the unique message such that 
\begin{equation}
(u^n(l),y^n)\in\calT_1\label{eqn:packing}
\end{equation}
for some $u^n(l)\in\calC(\hatm)$. If there is no such unique $\hatm$ declare an error. 

{\em Analysis of Error Probability}: Assume that $m=1$ and $L$ denotes the random index chosen by the encoder. Note that $L=L(S^n)$ is a random function of the random state sequence $S^n$. We will denote the chosen codeword interchangeably by $U^n(L)$ or $F_{\calC}(S^n) \in\calU^n$. The latter notation makes it clear that the chosen codeword is a random function of the state. The randomness of $F_{\calC}(S^n)$ comes not only from the random state $S^n$ but also from the random codewords  in $\calC$. There are three sources of error:
\begin{align}
\calE_1 &:= \{ \forall \,U^n(l)\in \calC(1):(U^n(l),S^n) \notin\calA \}\\
\calE_2 &:= \{  (U^n(L),Y^n) \notin\calT_1 \}\\
\calE_3 &:= \{  \exists \, U^n(\till) \notin\calC(1): (U^n(\till),Y^n) \in\calT_1 \} \label{eqn:calE_3}
\end{align}
In term of $\calE_1,\calE_2$ and $\calE_3$, the probability of error $\bbP(\calE)$ defined in~\eqref{eqn:error_av} can be bounded as 
\begin{equation}
\bbP(\calE)\le \bbP(\calE_1)+\bbP(\calE_2\cap\calE_1^c)+\bbP(\calE_3) . \label{eqn:sum_prob}
\end{equation}
We bound each of the probabilities above: First consider $\bbP(\calE_1 )$:
\begin{align} 
\bbP(\calE_1) &=\bbP( \forall \,U^n(l)\in \calC(1):(U^n(l),S^n) \notin\calA ) \\*
&=\sum_{s^n}P_{S^n}(s^n) \left [ \sum_{u^n: (u^n,s^n)\notin\calA} P_{U^n}(u^n)\right]^{|\calC(1)|}  ,\label{eqn:indepu}
\end{align}
where~\eqref{eqn:indepu} holds because the codewords  $U^n(l)$ are generated independently of each other and they are independent of the state sequence $S^n$.  We now upper bound~\eqref{eqn:indepu} as follows:
\begin{align} 
\bbP(\calE_1)\le\sum_{s^n }P_{S^n}(s^n) \left[ 1-\sum_{u^n} P_{U^n}(u^n)\chi(u^n  ,s^n)\right]^{|\calC(1)|} \label{eqn:oneminus}
\end{align}
where $\chi(u^n  ,s^n)$ is the indicator of the set $\calA\cap\calT_2$, i.e.,
\begin{equation}
 \chi(u^n  ,s^n):=\bone\{ (u^n,s^n)\in\calA\cap\calT_2\}.  \label{eqn:defJg}
\end{equation}
Clearly, by using the definition of $\calT_2$ in~\eqref{eqn:calT2}, we have that for all $(u^n,s^n)\in\calA\cap\calT_2$,
\begin{equation}
P_{U^n}(u^n)\ge P_{U^n|S^n}(u^n|s^n)\exp( - n (\overI(\bU;\bS)+ \gamma_2 )). \label{eqn:tilt} 
\end{equation}
Thus substituting the bound in~\eqref{eqn:tilt} into~\eqref{eqn:oneminus}, we have 
\begin{align} 
&\bbP(\calE_1) \nn\\*
 &\le\sum_{s^n }P_{S^n}(s^n) \bigg[ 1-\exp( - n (\overI(\bU;\bS)+ \gamma_2 ))\nn\\*
&\quad \times  \sum_{u^n} P_{U^n|S^n}(u^n|s^n)\chi(u^n  ,s^n)\bigg]^{|\calC(1)|} \\
&\le\sum_{s^n}P_{S^n}(s^n) \Bigg[ 1-\sum_{u^n} P_{U^n|S^n}(u^n|s^n)\chi(u^n  ,s^n) \nn\\
&\quad     + \exp \left[ -\exp( - n (\overI(\bU;\bS)+ \gamma_2 )) |\calC(1)|\right]\Bigg] \label{eqn:ineq} ,
\end{align}
where~\eqref{eqn:ineq} comes from the   inequality $(1-xy)^k\le 1-x+\exp(yk)$. Recall that the size of the subcodebook $\calC(1)$ is $|\calC(1)|=\lceil\exp( n ( \overI(\bU;\bS)+2\gamma_2))\rceil$. Thus, 
\begin{align} 
&\bbP(\calE_1 )\nn\\*
&\le\sum_{s^n}P_{S^n}(s^n) \bigg[ 1-\sum_{u^n} P_{U^n|S^n}(u^n|s^n)\chi(u^n  ,s^n)  \nn\\*
&\qquad + \exp \left( -\exp( n  \gamma_2  ) \right)\bigg] \label{eqn:ineq2}  \\
&= \bbP \left( (U^n,S^n)\in\calA^c\cup\calT_2^c \right) + \exp \left( -\exp( n  \gamma_2  ) \right)   \label{eqn:defJ}\\
&\le\bbP \left( (U^n,S^n)\in\calA^c \right)+\bbP( (U^n,S^n)\in\calT_2^c)  \nn\\*
&\qquad+ \exp \left( -\exp( n   \gamma_2 ) \right)  \label{eqn:ub}\\
&=\bbP \left( (U^n,S^n)\in\calA^c \right)+\pi_2+ \exp \left( -\exp( n  \gamma_2 ) \right)  \label{eqn:ub3} ,
\end{align}
where~\eqref{eqn:defJ} follows from the definition of $\chi(u^n  ,s^n)$ in~\eqref{eqn:defJg}, and~\eqref{eqn:ub3} follows from   the  definition of $\pi_2$ in~\eqref{eqn:defpi2}.  We now bound  the first term in~\eqref{eqn:ub3}. We have 
\begin{align}
&\bbP\left( (U^n,S^n)\in\calA^c \right)  \nn\\
&= \bbP\left(\eta(U^n,S^n)>\pi_1^{1/2} \right)\\
& \le \pi_1^{-1/2}\bbE\left(\eta(U^n,S^n)\right) \label{eqn:marko}\\
&= \pi_1^{-1/2}\sum_{u^n,s^n}P_{U^n,S^n} (u^n,s^n)\eta(u^n,s^n)\\
&= \pi_1^{-1/2}\sum_{u^n,s^n}P_{U^n,S^n} (u^n,s^n)  \nn\\*
&\quad \times  \sum_{x^n}\sum_{y^n: (u^n, y^n)\notin\calT_1}W^n(y^n|x^n,s^n) P_{X^n|U^n,S^n}(x^n|u^n,s^n)\label{eqn:def_eta}\\
&= \pi_1^{-1/2}\sum_{\substack{u^n,s^n,x^n,y^n:\\(u^n,y^n)\notin\calT_1}}P_{U^n,S^n,X^n,Y^n} (u^n,s^n,x^n,y^n)  \label{eqn:markov_ch}\\
&= \pi_1^{-1/2}\bbP( (U^n, Y^n)\notin\calT_1)\\
& \le \pi_1^{ 1/2} \label{eqn:def_T1}
\end{align}
where~\eqref{eqn:marko} is by Markov's inequality  and~\eqref{eqn:def_eta} is due to the definition of  $\eta(u^n, s^n)$ in~\eqref{eqn:eta_def}. Equality~\eqref{eqn:markov_ch} follows by the Markov chain $U^n-(X^n,S^n)-Y^n$  and~\eqref{eqn:def_T1} is by the definition of $\pi_1$ in~\eqref{eqn:defpi1}.  Combining the bounds in~\eqref{eqn:ub3} and~\eqref{eqn:def_T1} yields
\begin{equation}
\bbP(\calE_1 )\le  \pi_1^{ 1/2} +\pi_2 +  \exp \left( -\exp( n   \gamma_2 ) \right)   \label{eqn:boundE1} .
\end{equation}

Now we bound $\bbP(\calE_2\cap\calE_1^c)$. Recall that the mapping from the noncausal state sequence $S^n$ to the chosen codeword $U^n(L)$ is denoted as $F_{\calC}(S^n)$. Define the event
\begin{equation}
\calF := \{ (F_{\calC}(S^n),S^n)\in\calA\}  . \label{eqn:defcalG}
\end{equation}
Then, $\bbP(\calE_2\cap\calE_1^c)$ can be bounded as 
\begin{align}
\bbP(\calE_2\cap\calE_1^c) &=  \bbP(\calE_2\cap\calE_1^c\cap\calF)+\bbP(\calE_2\cap\calE_1^c\cap\calF^c )\\
&\le\bbP(\calE_2\cap\calF)+\bbP(\calE_1^c\cap\calF^c ) . \label{eqn:split}
\end{align}
The second term in~\eqref{eqn:split} is  identically zero because given that $\calE_1^c$ occurs, we can successfully find a $u^n\in\calC(1)$ such that $(u^n,s^n)\in\calA$ hence $\calE_1^c\cap\calF^c=\emptyset$.   Refer to  the encoding step in~\eqref{eqn:covering}.
So  we consider the first term in~\eqref{eqn:split}. Let $\bbE_{\calC}[\fndot ]$ be the expectation over the random codebook $\calC$, i.e., $\bbE_{\calC}[ \phi(\calC)] =\sum_{\frakc}\bbP(\calC=\frakc)\phi(\frakc)$, where $\frakc$ runs over all possible  sets of sequences $\{u^n(l ) \in\calU^n:l\in [1:\exp(n\tilR)]\}$.  Consider,
\begin{align}
&\bbP(  \{(F_{\calC}(S^n), Y^n)\notin\calT_1 \} \cap\{(F_{\calC}(S^n),S^n)\in\calA\} ) \nn\\
&= \bbE_{\calC}\Bigg[\sum_{ \substack{(s^n, y^n):( F_{\calC}(s^n), y^n) \notin\calT_1 \\ (F_{\calC}(s^n),s^n)\in\calA}} P_{S^n}(s^n)   \nn\\*
& \qquad\times \sum_{x^n}W^n(y^n|x^n,s^n)P_{X^n|U^n,S^n} (x^n|F_{\calC}(s^n),s^n)  \Bigg] \label{eqn:fixed_u} \\
&= \bbE_{\calC}\Bigg[\sum_{ s^n:   (F_{\calC}(s^n),s^n)\in\calA} P_{S^n}(s^n)  \nn\\*
& \qquad\times \bigg(\sum_{x^n} \sum_{y^n: ( F_{\calC}(s^n), y^n) \notin\calT_1 }  W^n(y^n|x^n,s^n) \nn\\*
&\qquad\times P_{X^n|U^n,S^n} (x^n|F_{\calC}(s^n),s^n)  \bigg) \Bigg].  \label{eqn:rearrange}  
\end{align}
Equality~\eqref{eqn:fixed_u} can be explained as follows: Conditioned on  $\{\calC=\frakc\}$ for some (deterministic) codebook $\frakc$, the mapping $F_{\frakc}:\calS^n\to\calU^n$  is   deterministic  (cf. the encoding step in~\eqref{eqn:covering}) and  thus $\{U^n=F_\frakc(s^n)\}$   holds if  $\{S^n=s^n\}$ holds. Therefore,   the conditional distribution of $Y^n$ given  $\{S^n=s^n\}$    and hence also $\{U^n=F_{\frakc}(s^n)\}$   is $\sum_{x^n}   W^n(\fndot|x^n,s^n) P_{X^n|U^n,S^n} (x^n|F_{\frakc}(s^n),s^n)$ for fixed $\frakc$.  The step~\eqref{eqn:fixed_u} differs subtly from  the proofs of the general Wyner-Ziv problem~\cite{iwata02} and the general lossless source coding with coded side information problem~\cite{ miyake, Kuz12}. In~\cite{iwata02}, \cite{miyake} and~\cite{Kuz12}, the equivalent of $Y^n$ just depends only  implicitly the auxiliary $U^n$ through another variable, say\footnote{In the Wyner-Ziv problem~\cite{iwata02}, $\tilX^n$ is the source to be reconstructed to within some distortion with the help of $Y^n$ and in the WAK problem~\cite{ miyake,Kuz12},   $Y^n$ is the source to be almost-losslessly  transmitted with the help of a coded version of $\tilX^n$.}  $\tilX^n$ (i.e.,  $Y^n-\tilX^n-U^n$ forms a Markov chain). In the Gel'fand-Pinsker problem, $Y^n$ depends on $X^n$ and $S^n$, the former being a (stochastic) function of both $U^n$ and $S^n$. Thus, given $\{S^n=s^n\}$, $Y^n$ also depends     on the state $S^n$ and the auxiliary $U^n$ through the   codebook  $\calC$ and the covering procedure specified by $F_{\calC}$.  Now using the definitions of $\eta(u^n, s^n)$ and $\calA$ in~\eqref{eqn:eta_def} and~\eqref{eqn:defcalA} respectively, we can bound \eqref{eqn:rearrange} as follows: 
\begin{align}
&\bbP(  \{(F_{\calC}(S^n), Y^n)\notin\calT_1 \} \cap\{(F_{\calC}(S^n),S^n)\in\calA\} ) \nn\\
&= \bbE_{\calC}\left[\sum_{ s^n:   (F_{\calC}(s^n),s^n)\in\calA} P_{S^n}(s^n)\eta(  F_{\calC}(s^n),s^n) \right]  \label{eqn:def_eta2} \\
&\le \bbE_{\calC}\left[\sum_{ s^n:   (F_{\calC}(s^n),s^n)\in\calA} P_{S^n}(s^n)\pi_1^{1/2}  \right] \label{eqn:boundE2}  \\ 
&\le \pi_1^{1/2} . \label{eqn:boundE20}  
\end{align}
Uniting~\eqref{eqn:split} and~\eqref{eqn:boundE20} yields 
\begin{equation}
\bbP(\calE_2\cap\calE_1^c)\le \pi_1^{1/2}. \label{eqn:boundPE2}
\end{equation}

Finally, we consider the probability $\bbP(\calE_3)$:
\begin{align}
\bbP(\calE_3) &=\bbP \left( (U^n(\till),Y^n) \in\calT_1 \, \mbox{  for some  } \, U^n(\till)  \notin\calC(1)\right) \\
&\le\sum_{\till= \lceil\exp(n (\tilR-R)) \rceil+1}^{\lceil\exp(n\tilR)\rceil}  \bbP \left(    (U^n(\till),Y^n) \in\calT_1  \right) , \label{eqn:ub2} 
\end{align}
where \eqref{eqn:ub2}   follows from the union bound and the fact that the indices of the confounding codewords belong to the   set $[ \exp(n (\tilR-R))+1:\exp(n\tilR)]$. Now, we upper bound the probability in~\eqref{eqn:ub2}. Note that if $\till \in[ \exp(n (\tilR-R))+1:\exp(n\tilR)]$, then $U^n(\till)$ is independent of $Y^n$. Thus, 
\begin{align}
& \bbP \left(    (U^n(\till),Y^n) \in\calT_1  \right) \nn\\
 &=\sum_{(u^n,y^n)\in\calT_1}P_{U^n}(u^n)P_{Y^n}(y^n) \\
 &\le\sum_{(u^n,y^n)\in\calT_1}P_{U^n}(u^n)P_{Y^n|U^n}(y^n|u^n) \nn\\* 
 &\qquad\times  \exp \left(-n (\underI(\bU;\bY)- \gamma_1)\right)  \label{eqn:defT1}\\
 &\le \exp \left(-n (\underI(\bU;\bY)- \gamma_1)\right)  \label{eqn:sum1}
 \end{align}
where~\eqref{eqn:defT1} follows from the definition of $\calT_1$ in~\eqref{eqn:calT1}. Now, substituting~\eqref{eqn:sum1} into~\eqref{eqn:ub2} yields
\begin{align}
\bbP(\calE_3) &\le \exp(n\tilR)\exp \left(-n (\underI(\bU;\bY)- \gamma_1)\right)  \\
 &= \exp(n (\underI(\bU;\bY)-2\gamma_1))  \nn\\*
 &\qquad\times \exp \left(-n (\underI(\bU;\bY)- \gamma_1)\right)  \label{eqn:boundE3a}\\
& = \exp \left(-n \gamma_1 \right)   \label{eqn:boundE3}
\end{align}
where~\eqref{eqn:boundE3a} follows from the definition of $\tilR$ in~\eqref{eqn:iflate}. Uniting~\eqref{eqn:sum_prob},~\eqref{eqn:boundE1},~\eqref{eqn:boundPE2} and~\eqref{eqn:boundE3}  shows that $\bbP(\calE)\le\rho_n$, where $\rho_n$ is defined in~\eqref{eqn:bound_err}. By the selection lemma, there exists a (deterministic) code whose average error probability is no larger than $\rho_n$.  \end{proof}

\subsection{Proof of Corollary~\ref{cor:both}}\label{app:both}
\begin{proof} 
In this problem where the state information is available at the encoder and the decoder, we   can regard the output $\bY$ of the Gel'fand-Pinsker problem as  $(\bY,\bS)$. For achievability, we lower bound the generalization of the Cover-Chiang~\cite{coverChiang} result in~\eqref{eqn:gp_res_cc} with $\bS_{\mathrm{d}}=\bS_{\mathrm{e}}=\bS$ as follows:
\begin{align}
C_{\mathrm{CC}} &=\sup_{ \bU-(\bX,\bS)-\bY}\underI(\bU;\bY,\bS)-\overI(\bU;\bS) \label{eqn:YS}\\
&\ge \underI(\bX^*;\bY^*,\bS)-\overI(\bX^*;\bS) \label{eqn:x_star}\\
&= \underI(\bX^*;\bY^*,\bS)-\underI(\bX^*;\bS) \label{eqn:use_assump}\\
&\ge \underI(\bX^*;\bY^*|\bS) , \label{eqn:prop_pliminf}
\end{align}
where~\eqref{eqn:x_star} follows because the choice ($\{P_{X^n|S^n}^*\}_{n=1}^{\infty} , U^n=\emptyset)$ belongs to the constraint set in~\eqref{eqn:YS}, \eqref{eqn:use_assump} uses the assumption in~\eqref{eqn:both} and \eqref{eqn:prop_pliminf} follows from the basic information spectrum inequality  $\pliminf_{n\to\infty}(A_n+B_n)\le\pliminf_{n\to\infty} A_n+\plimsup_{n\to\infty}B_n$~\cite[pp.~15]{Han10}. This shows that $C_{\mathrm{ED}} = \underI(\bX^*;\bY^*|\bS)$ is an achievable rate. 

For the converse, we upper bound~\eqref{eqn:YS} as follows:
\begin{align}
C_{\mathrm{CC}}&\le \sup_{ \bU-(\bX,\bS)-\bY}\underI(\bU;\bY|\bS) \label{eqn:prop_plim}\\
&\le \sup_{ \bU-(\bX,\bS)-\bY}\underI(\bX;\bY|\bS) \label{eqn:data_proc}\\
&= \sup_{  \bX }\underI(\bX;\bY|\bS) \label{eqn:noU}
\end{align}
where \eqref{eqn:prop_plim} follows from the superadditivity of $\pliminf$,  \eqref{eqn:data_proc} follows from the $\pliminf$-version of the (conditional) data processing inequality~\cite[Theorem 9]{VH94} and~\eqref{eqn:noU} follows because there is no $\bU$ in the objective function in~\eqref{eqn:data_proc} so taking $U^n=\emptyset$ does not violate optimality in~\eqref{eqn:data_proc}. Since~\eqref{eqn:noU} implies that all achievable rates are bounded above by  $C_{\mathrm{ED}}$, this proves the converse. 
\end{proof}
\subsection{Proof of Corollary~\ref{cor:mem}  }\label{app:converse}
\begin{proof}
For achievability, note that since the channel and state are memoryless, we lower bound~\eqref{eqn:gp_res}  by replacing the constraint set  with the set of  conditional distributions of the product from, i.e., $P_{X^n,U^n|S^n} =\prod_{i=1}^n P_{X_i,U_i|S_i}$  for some $\{P_{X_i,U_i|S_i}:\calS\to\calX\times\calU\}_{i=1}^n$. Let $\calM_{\bW,\bS}$ (which stands for ``memoryless'') be the set of all $(\bU,\bX,\bS,\bY)$ such that $P_{X^n,U^n|S^n}=\prod_{i=1}^n P_{X_i,U_i|S_i}$ for some $\{P_{X_i,U_i|S_i}\}_{i=1}^n$, the channels coincide with $\{\prod_{i=1}^n W_i\}_{n=1}^{\infty}$ and the states coincide with $\{\prod_{i=1}^n P_{S_i}\}_{n=1}^{\infty}$. Let $(S_i, U_i^*, X_i^*, Y_i^*)$ be distributed according to $P_{S_i}\circ P_{X_i, U_i|S_i}^*\circ W_i$,  the optimal distribution in~\eqref{eqn:ith_opt}. Consider,
\begin{align}
C &\ge \sup_{({\bU}, {\bX}, {\bS}, {\bY})\in\calM_{\bW,\bS}}  \bigg\{ \pliminf_{n\to\infty}\frac{1}{n}\log\frac{ P_{Y^n|U^n}(Y^n|U^n)}{P_{Y^n}(Y^n)}   \nn\\* &\qquad -\plimsup_{n\to\infty}\frac{1}{n}\log\frac{ P_{U^n|S^n}(U^n|S^n)}{P_{U^n}(U^n)}   \bigg\} \label{eqn:memoryless}\\
&\ge \pliminf_{n\to\infty}\frac{1}{n}\log\frac{ P_{(Y^n)^*|(U^n)^*}((Y^n)^*|(U^n)^*)}{P_{(Y^n)^*}( (Y^n)^* )}  \nn\\* 
&\qquad -  \plimsup_{n\to\infty}\frac{1}{n}\log\frac{ P_{(U^n)^*|(S^n)^*}((U^n)^*|(S^n)^*)}{P_{(U^n)^*}((U^n)^*)}  \label{eqn:memoryless1}\\
&= \liminf_{n\to\infty}\frac{1}{n}\sum_{i=1}^n  \bbE\left[ \log \frac{P_{Y_i^*|U_i^*}(Y_i^*|U_i^*)}{P_{Y_i^*}(Y_i^*)} \right]  \nn\\*  
&\qquad -  \limsup_{n\to\infty}\frac{1}{n}\sum_{i=1}^n  \bbE\left[ \log \frac{P_{U_i^*|S_i}(U_i^*|S_i )}{P_{U_i^*}(U_i^*)} \right]  \label{eqn:chey} \\
&=  \liminf_{n\to\infty}\frac{1}{n}\sum_{i=1}^n I(U_i^*;Y_i^*)  -  \limsup_{n\to\infty}\frac{1}{n}\sum_{i=1}^n  I(U_i^*;S_i) \label{eqn:defmi}
\end{align}
where~\eqref{eqn:memoryless1}  follows by substituting the optimal $(S_i, U_i^*, X_i^*, Y_i^*)$  into~\eqref{eqn:memoryless}. Inequality~\eqref{eqn:chey}  follows from   Chebyshev's inequality where we used the  memoryless  assumption and the assumption that the alphabets are finite so the variances of the information densities are uniformly bounded~\cite[Remark~3.1.1]{Han10}. Essentially the limit inferior (resp.\ superior) in probability of the normalized information densities  become the regular limit inferior (resp.\ superior) of averages of mutual informations under the memoryless assumption. Now, we assume   the  first limit in~\eqref{eqn:lims_exist} exists. Then, the $\limsup$ in \eqref{eqn:defmi} is in fact a limit and we have 
\begin{align}
C &\ge  \liminf_{n\to\infty}\frac{1}{n}\sum_{i=1}^n I(U_i^*;Y_i^*)  -  \lim_{n\to\infty}\frac{1}{n}\sum_{i=1}^n  I(U_i^*;S_i) \label{eqn:be_limis} \\
&=  \liminf_{n\to\infty}\frac{1}{n}\sum_{i=1}^n I(U_i^*;Y_i^*)  - I(U_i^*;S_i) \label{eqn:super-add} 
\end{align}
where~\eqref{eqn:super-add} follows from the fact that $\liminf_{n\to\infty}(a_n+b_n) = \liminf_{n\to\infty} a_n+\lim_{n\to\infty}b_n$ if $b_n$ converges.    If instead the second limit in \eqref{eqn:lims_exist} exists, the argument  from~\eqref{eqn:defmi} to \eqref{eqn:super-add} proceeds along exactly the same lines.  The proof of the direct part is completed by invoking the definition of $C(W_i,P_{S_i})$ in \eqref{eqn:ith_opt}.


For the converse,  we    {\em only} assume that $|\calS|<\infty$. Let $(\hatS^n,\hatU^n,\hatX^n, \hatY^n)$ be  dummy variables distributed as  $P_{S^n}\circ P_{U^n|S^n} \circ P_{X^n|U^n,S^n}\circ W^n$. Note  that $P_{S^n}$ and $W^n$ are assumed to be memoryless.  From~\cite[Theorem~3.5.2]{Han10},
\begin{align}
\underI(\hat{\bU};\hat{\bY}) & \le \liminf_{n\to\infty} \frac{1}{n}I(\hatU^n;\hatY^n) \label{eqn:no_req} 
\end{align}
and if $|\calS|<\infty$, we also have 
\begin{align}
\overI(\hat{\bU};\hat{\bS}) &\ge \limsup_{n\to\infty} \frac{1}{n}I(\hatU^n;\hatS^n) . \label{eqn:req_finite}
\end{align}
Now we can upper bound the objective function in~\eqref{eqn:gp_res} as follows:
\begin{align}
&\underI(\hat{\bU};\hat{\bY})-\overI(\hat{\bU};\hat{\bS}) \nn\\
&\le\liminf_{n\to\infty} \frac{1}{n}I(\hatU^n;\hatY^n) + \liminf_{n\to\infty} -\, \frac{1}{n}I(\hatU^n;\hatS^n) \\
&\le\liminf_{n\to\infty} \frac{1}{n} \left[I(\hatU^n;\hatY^n) - I(\hatU^n;\hatS^n)\right], \label{eqn:liminf_bound}
\end{align}
where~\eqref{eqn:liminf_bound} follows from the superadditivity of the limit inferior. Hence, it suffices to single-letterize the expression in $[\fndot]$ in~\eqref{eqn:liminf_bound}.   To start, consider 
\begin{align}
&I(\hatU^n;\hatY^n) - I(\hatU^n;\hatS^n) \nn\\
&=\sum_{i=1}^n I(\hatU^n;\hatY_i|\hatY^{i-1}, \hatS_{i+1}^n)-I(\hatU^n;\hatS_i|\hatY^{i-1}, \hatS_{i+1}^n) \label{eqn:c_sum}\\
&=\sum_{i=1}^n I(\hatU^n, \hatY^{i-1}, \hatS_{i+1}^n ;\hatY_i)- I(\hatY_i;\hatY^{i-1}, \hatS_{i+1}^n)   \nn\\
&\qquad  -I(\hatU^n, \hatY^{i-1}, \hatS_{i+1}^n;\hatS_i)+ I(\hatS_i;\hatY^{i-1}, \hatS_{i+1}^n) ,\label{eqn:chain_rule}
\end{align}
where~\eqref{eqn:c_sum} follows from the key identity in Csisz\'ar and K\"orner~\cite[Lemma~17.12]{Csi97} and~\eqref{eqn:chain_rule} follows from the chain rule for mutual information. Now, we relate  the (sum of the) second term to the (sum of the)  fourth term:
\begin{align}
\sum_{i=1}^n I(\hatY_i;\hatY^{i-1}, \hatS_{i+1}^n) &\ge\sum_{i=1}^n I(\hatY_i; \hatS_{i+1}^n|\hatY^{i-1})\label{eqn:mi_nneg}\\
&=\sum_{i=1}^n I(\hatS_i; \hatY^{i-1}|\hatS_{i+1}^n) \label{eqn:c_sum1}\\
&=\sum_{i=1}^n I(\hatS_i; \hatY^{i-1},\hatS_{i+1}^n) , \label{eqn:indep_S} 
\end{align}
where ~\eqref{eqn:c_sum1} follows from  the Csisz\'ar-sum-identity~\cite[Chapter~2]{elgamal} and~\eqref{eqn:indep_S} follows because $\hatS^n=(\hatS_1,\ldots, \hatS_n)$ is a memoryless process.   Putting together~\eqref{eqn:liminf_bound}, \eqref{eqn:chain_rule} and \eqref{eqn:indep_S}, we have the upper bound:
\begin{align}
C\le\liminf_{n\to\infty}\frac{1}{n}\sum_{i=1}^n & I(\hatU^n, \hatY^{i-1}, \hatS_{i+1}^n;\hatY_i) \nn\\*
 &-I(\hatU^n, \hatY^{i-1}, \hatS_{i+1}^n;\hatS_i). \label{eqn:single_letter_ub}
\end{align}
Note that~\eqref{eqn:single_letter_ub} holds for all $P_{X^n,U^n|S^n}$ of the product form (i.e., $(\hat{\bU},\hat{\bX}, \hat{\bS}, \hat{\bY})\in \calM_{\bW,\bS}$) because the state and channel are memoryless.  Now  define $U_i:=(\hatU^n, \hatY^{i-1}, \hatS_{i+1}^n)$ and $Y_i:=\hatY_i, X_i:=\hatX_i$ and $S_i:=\hatS_i$. Clearly, the Markov chain $ U_i-(X_i,S_i)-Y_i$ is satisfied for all $i\in [1:n]$. Substituting these identifications  into~\eqref{eqn:single_letter_ub} yields
\begin{align}
C&\le \liminf_{n\to\infty}\frac{1}{n}\sum_{i=1}^n\max_{ U_i-( X_i,  S_i)- Y_i}I( U_i; Y_i)-I(U_i;S_i),
\end{align}
which upon invoking the definition of $C(W_i,P_{S_i})$ in \eqref{eqn:ith_opt} completes the converse proof of Corollary~\ref{cor:mem}.

 We remark that in the single-letterization procedure for the converse, it seems as if we did not use the assumption that the transmitted message is independent of the state. This is not true because in the converse proof of Theorem~\ref{thm:gp_ach} we did  in fact  use this key assumption. See~\eqref{eqn:indepUS} and~\eqref{eqn:useindepUS} where the message is represented by $\bU$.   \end{proof}

\subsection{Verification of \eqref{eqn:halfhalf} in Example \ref{eg:nonerg}}\label{app:example} 
Let $\calE_n:=\calE\cap[1:n]$ and $\calO_n:=\calO\cap[1:n]$  be respectively, the set of even and odd integers up to some $n\in\bbN$.  To verify~\eqref{eqn:halfhalf}, we use the result in~\eqref{eqn:mless} and the definitions of the channel and state in \eqref{eqn:ch_eg} and \eqref{eqn:st_eg} respectively. We first split the sum into odd and even parts as follows:
\begin{align} \label{eqn:Mless}
C_{\mathrm{M'less}} =\frac{1}{2}\liminf_{n\to\infty} \bigg[\frac{2}{n} &  \sum_{i\in\calO_n} C(W_i, P_{S_i}) \nn\\*
 &+\frac{2}{n}\sum_{ i\in\calE_n}  C(W_i, P_{S_i})  \bigg]
\end{align}
For the even part, each of the summands is    $C(\tilW_{\mathrm{c}},Q_{{\mathrm{b}}})$ so the sequence $b_n:=\frac{2}{n}\sum_{  i\in\calE_n}  C(W_i, P_{S_i})$   converges and the limit is $C(\tilW_{\mathrm{c}},Q_{{\mathrm{b}}})$ .  Let $a_n:=\frac{2}{n} \sum_{ i\in\calO_n} C(W_i, P_{S_i})$ be the odd part in~\eqref{eqn:Mless}. A basic fact in analysis  states that $\liminf_{n\to\infty}(a_n+b_n)= \liminf_{n\to\infty} a_n+\lim_{n\to\infty}b_n$ if $b_n$ has a limit. Hence, the above $\liminf$ is 
\begin{align} 
C_{\mathrm{M'less}} &=\frac{1}{2}\liminf_{n\to\infty} \left[\frac{2}{n} \sum_{  i\in\calO_n} C(W_i, P_{S_i})\right]+\frac{1}{2}C(\tilW_{\mathrm{c}},Q_{{\mathrm{b}}})   \label{eqn:apply_limits0} \\
&=\frac{1}{2}\liminf_{n\to\infty} \bigg[\frac{2}{n}  \left| \calO_n\cap\calJ \right| C(\tilW_{\rma}, Q_{{\mathrm{a}}}) \nn\\*
&\quad +\frac{2}{n} \left| \calO_n\cap\calJ^c \right| C(\tilW_{\rmb}, Q_{{\mathrm{a}}})\bigg]+\frac{1}{2}C(\tilW_{\mathrm{c}},Q_{{\mathrm{b}}}).  \label{eqn:apply_limits}
\end{align}
It can be verified that $\liminf_{n\to\infty} \frac{2}{n}|\calO_n\cap\calJ|=\frac{1}{3}$ and   $\limsup_{n\to\infty} \frac{2}{n}|\calO_n\cap\calJ|=\frac{2}{3}$. Thus, the $\liminf$ in~\eqref{eqn:apply_limits} is  
\begin{align}
\liminf_{n\to\infty}a_n&= \frac{2}{3}\min \left\{C(\tilW_{\mathrm{a}},Q_{\mathrm{a}}), C(\tilW_{\mathrm{b}},Q_{\mathrm{a}})\right\}  \nn\\*
&\qquad + \frac{1}{3}\max \left\{ C(\tilW_{\mathrm{a}},Q_{\mathrm{a}}), C(\tilW_{\mathrm{b}},Q_{\mathrm{a}})\right\} .
\end{align}
which, by definition, is equal to $G(Q_{\mathrm{a}})$ in \eqref{eqn:CQ}. This completes the verification of \eqref{eqn:halfhalf}.


\subsection{Proof of Corollary~\ref{cor:mixed1}} \label{sec:prf:cor1}
\begin{proof}
Fix $P_{X^n,U^n|S^n}$. The key observation  is to note that the joint distribution of $(S^n,U^n,X^n,Y^n)$ can be written as a convex combination of the distributions in~\eqref{eqn:product_mixed}
\begin{align}
&P_{S^n,U^n,X^n,Y^n}(s^n,u^n,x^n,y^n)  \nn\\
&= \left[ \sum_{l=1}^{\infty}\beta_l P_{S_l^n}(s^n) \right]P_{X^n, U^n|S^n}(u^n,x^n|s^n) \nn\\*
&\qquad \qquad\times \left[\sum_{k=1}^{\infty}\alpha_k W_k^n(y^n|x^n, s^n)\right] \\
 &=\sum_{k=1}^{\infty}\sum_{l=1}^{\infty}\alpha_k\beta_l \Big\{  P_{S_l^n}(s^n) P_{X^n, U^n|S^n}(u^n,x^n|s^n)   \nn\\*
 &\qquad \qquad\times   W_k^n(y^n|x^n, s^n)\Big\}\\
  &=\sum_{k=1}^{\infty}\sum_{l=1}^{\infty}\alpha_k\beta_l P_{S^n_{l},U^n_{l}, X^n_{l}, Y^n_{kl}}(s^n,u^n,x^n,y^n) , \label{eqn:expand_mix}
\end{align}
where~\eqref{eqn:expand_mix} follows from the definition of $P_{S^n_{l},U^n_{l}, X^n_{l}, Y^n_{kl}}$ in~\eqref{eqn:product_mixed}.   By Tonelli's theorem, the marginals of $(U^n,Y^n)$  and $(U^n,S^n)$ are   given  respectively by
\begin{align}
P_{U^n ,Y^n}(u^n ,y^n) &=\sum_{k=1}^{\infty}\sum_{l=1}^{\infty}\alpha_k\beta_l P_{ U^n_{l} , Y^n_{kl}}(u^n,y^n)  \label{eqn:marg_countable}\\
P_{U^n ,S^n}(u^n ,s^n) &=\sum_{l=1}^{\infty} \beta_l P_{ U^n_{l} , S_l^n}(u^n,s^n),  \label{eqn:marg_countable2}
\end{align}
where in~\eqref{eqn:marg_countable2} we used the fact that $\sum_{k=1}^{\infty}\alpha_k=1$. The number of terms in the  sums in~\eqref{eqn:marg_countable} and~\eqref{eqn:marg_countable2} are countable.  By Lemma~3.3.2 in~\cite{Han10} (the spectral inf-mutual information rate of a mixed source is the infimum of the constituent  spectral inf-mutual information rates of those processes with positive weights), we know that 
\begin{equation}
\underI(\bU;\bY)=\inf_{(k,l)\in\calK\times\calL}\underI(\bU_{kl};\bY_{kl}),
\end{equation}
where recall that $\calK =\{k\in\bbN:\alpha_k>0\}$ and $\calL =\{l\in\bbN:\beta_l>0\}$. Analogously,   
\begin{equation}
\overI(\bU;\bS)=\sup_{l\in \calL}\overI(\bU_{l};\bS_{l}). 
\end{equation}
This completes the proof  of~\eqref{eqn:gp_res_mixed}.

The achievability statement in \eqref{eqn:gp_res_mixed_st} follows by considering the optimal $P_{X,U|S}$  (in the chain $U-(X,S)-Y$) and  defining the i.i.d.\  random variables $(S_l^n,X_l^n,U_l^n, Y_{kl}^n)\sim\prod_{i=1}^n P_{S_l}(s_i)   P_{X,U|S}(x_i,u_i|s_i)W_k (y_i|x_i,s_i)$.   Khintchine's law of large numbers~\cite[Lemma 1.3.2]{Han10} then asserts that for every $(k,l)\in\calK\times\calL$, we have $\underI(\bU_l;\bY_{kl})=I(U_l;Y_{kl})$ and $\overI(\bU_l;\bS_{l})=I(U_l;S_l)$, completing the proof of the lower bound in~\eqref{eqn:gp_res_mixed_st}.
\end{proof}


\subsection{Proof of Theorem~\ref{thm:coded}} \label{sec:prf:coded}
\begin{proof}
We will only sketch the proof for the direct part since it combines the proof of Theorem~\ref{thm:gp_ach} and the main result in Iwata and Muramatsu~\cite{iwata02} in a straightforward manner. In fact, this proof technique originated from Heegard and El Gamal~\cite{heegard}. Perform Wyner-Ziv coding for ``source'' $S^n$ at the state encoder with correlated ``side information'' $Y^n$. This generates a rate-limited description of $S^n$ at the decoder and the rate is approximately $\frac{1}{n}\log M_{\mathrm{d},n}$. Call this description $V^n \in\calV^n$. The rate constraint is given in~\eqref{eqn:wz} after going through the same steps as in~\cite{iwata02}. The description $V^n$  is  then used as common side information of the state (at the encoder and decoder) for the usual Gel'fand-Pinsker coding problem, resulting in~\eqref{eqn:cgp}. This part is simply a conditional version of Theorem~\ref{thm:gp_ach}.

For the converse part,
fix $\gamma>0$ and note that   
\begin{equation}
M_{\mathrm{d}, n}\le \exp(n (\Rd+\gamma)) \label{eqn:rate_d}
\end{equation}
 for $n$ large enough by the second inequality in~\eqref{eqn:defs_coded}. Thus,  if $V^n$ denotes an arbitrary random variable such that the cardinality of its support is no larger than $M_{\mathrm{d}, n}$, we can   check  that~\cite[Theorem 5.4.1]{Han10} 
\begin{align}
\bbP\left(\frac{1}{n}\log\frac{1}{P_{V^n}(V^n)}\ge\frac{1}{n}\log M_{\mathrm{d}, n}+\gamma \right)  \le\exp(-n\gamma). \label{eqn:compression_is}
\end{align}
We can further lower bound the left-hand-side of~\eqref{eqn:compression_is} using~\eqref{eqn:rate_d} which   yields
\begin{equation}
\bbP\left(\frac{1}{n}\log\frac{1}{P_{V^n}(V^n)}\ge\Rd+2\gamma \right)  \le \exp(-n\gamma). \label{eqn:subs_Md}
\end{equation}
Now consider,
\begin{align}
\Rd &\ge \overH(\bV) -2\gamma  \label{eqn:greaterH} \\
&\ge \overH(\bV)-\underH(\bV|\bS) -2\gamma   \label{eqn:nonneg_ent} \\
 & \ge\overI(\bV;\bS)-2\gamma \label{eqn:greaterI} \\
&\ge \overI(\bV;\bS)-\underI(\bV;\bY)-2\gamma . \label{eqn:pos_is}
\end{align}    
Inequality~\eqref{eqn:greaterH}  follows from~\eqref{eqn:subs_Md} and the definition of the spectral sup-entropy rate~\cite[Chapter~1]{Han10}.   Inequality~\eqref{eqn:nonneg_ent} holds because by \eqref{eqn:rate_d}, $V^n$ is supported on finitely many points in $\calV^n$ and so the spectral inf-conditional entropy rate  $\underH(\bV|\bS)$ is non-negative~\cite[Lemma~2(a)]{Ste96}. Inequality~\eqref{eqn:greaterI} follows because the limsup in probability  is  superadditive~\cite[Theorem~8(d)]{VH94}.  Finally, inequality~\eqref{eqn:pos_is} follows because the spectral inf-mutual information rate is non-negative~\cite[Theorem~8(c)]{VH94}. 





The upper bound on the transmission rate in~\eqref{eqn:cgp} follows by   considering a conditional version of the Verd\'{u}-Han converse. As in~\eqref{eqn:use_vh}, this yields the constraint
\begin{equation}
R\le\underI(\bU;\bY|\bV)+2\gamma, \label{eqn:conditional_vh}
\end{equation}  
where $\bU=\{U^n\}_{n=1}^{\infty}$ denotes a random variable representing the uniform choice of a message in $[1:M_n]$.    Note that $\bV=\{V^n\}_{n=1}^{\infty}$ is recoverable (available) at both encoder and decoder hence the conditioning on $\bV$ in~\eqref{eqn:conditional_vh}. Also note  that $(\bS,\bV)$ is independent of $\bU$ since $V^n$ is a function of $S^n$ and $S^n$ is independent of $U^n$. Hence $\overI(\bU;\bS,\bV)=0$. Because  
$
\overI(\bU;\bS|\bV)\le\overI(\bU;\bS,\bV)
$ 
and the  spectral sup-conditional mutual information rate is non-negative~\cite[Lemma 3.2.1]{Han10},  
\begin{equation}
\overI(\bU;\bS|\bV)=0.
\end{equation}
In view of~\eqref{eqn:conditional_vh}, we have
\begin{equation}
R\le\underI(\bU;\bY|\bV)-\overI(\bU;\bS|\bV)+2\gamma . \label{eqn:conditional_vh_minus}
\end{equation}
Since we have proved~\eqref{eqn:pos_is} and~\eqref{eqn:conditional_vh_minus}, we can now proceed in the same way we did for the unconditional case in the converse proof of Theorem~\ref{thm:gp_ach}. Refer to the steps~\eqref{eqn:useindepUS} to~\eqref{eqn:large_sup}.  Finally,    take $\gamma\to 0$ and this gives~\eqref{eqn:cgp} and~\eqref{eqn:wz}.
\end{proof}

\subsection{Proof of Corollary~\ref{cor:coded}}\label{app:coded_sm}

\begin{proof}
We only prove the converse since achievability follows easily   using i.i.d.\ random codes and Khintchine's law of large numbers~\cite[Lemma 1.3.2]{Han10}. Essentially, all four spectral inf- and spectral sup-mutual information rates in~\eqref{eqn:cgp} and~\eqref{eqn:wz} are mutual informations. For the converse, we fix $(\hat{\bU}, \hat{\bV})-(\hat{\bX},\hat{\bS})-\hat{\bY}$ and   lower bound $\Rd$ in~\eqref{eqn:wz} as follows:
\begin{align}
\Rd &\ge \overI(\hat{\bV};\hat{\bS})-\underI(\hat{\bV};\hat{\bY})\\
&\ge \limsup_{n\to\infty}\frac{1}{n}I(\hatV^n;\hatS^n)-\liminf_{n\to\infty}\frac{1}{n} I(\hatV^n;\hatY^n) \label{eqn:subpliminf}\\ 
&\ge\limsup_{n\to\infty}\frac{1}{n}  \left[ I(\hatV^n;\hatS^n) -I(\hatV^n;\hatY^n)\right]\\
&\ge\limsup_{n\to\infty}\frac{1}{n}  \bigg[  \sum_{i=1}^n I(\hatV^n, \hatY^{i-1}, \hatS_{i+1}^n;\hatS_i) \nn\\*
 &\qquad\qquad -I(\hatV^n, \hatY^{i-1}, \hatS_{i+1}^n;\hatY_i)\bigg]  ,\label{eqn:follow_steps}
\end{align}
where~\eqref{eqn:subpliminf} follows from the same reasoning as in~\eqref{eqn:no_req} and~\eqref{eqn:req_finite} (using the fact that $|\calS|<\infty$)   and~\eqref{eqn:follow_steps} follows the same steps as in the proof of Corollary~\ref{cor:stat}. In the same way, the condition on $R$ in~\eqref{eqn:cgp} can  be further upper bounded using the Csisz\'ar-sum-identity~\cite[Chapter 2]{elgamal} and memorylessness of $\hatS^n$  as follows
\begin{align}
R \le\liminf_{n\to\infty}\frac{1}{n}  \bigg[  \sum_{i=1}^n & I(\hatU^n ;\hatY_i|\hatV^n,\hatY^{i-1}, \hatS_{i+1}^n)  \nn\\*
&- I(\hatU^n ;\hatS_i|\hatV^n,\hatY^{i-1}, \hatS_{i+1}^n)\bigg] . \label{eqn:upper_bd_R}
\end{align}
From this point on, the  rest of the proof is standard. We let $V_i:=( \hatV^n, \hatY^{i-1}, \hatS_{i+1}^n)$ and $U_i:=( \hatU^n, V_i)$ and $Y_i, X_i$ and $S_i$ as in the proof of Corollary~\ref{cor:mem}. These identifications satisfy $(U_i, V_i)-(X_i, S_i)-Y_i$  and the proof can be completed as per the usual steps. \end{proof}

 \subsection*{Acknowledgements}
I am very grateful to Stark C.\ Draper for several discussions that led to the current work. This work is supported by NUS startup grants WBS R-263-
000-A98-750 (FoE) and WBS R-263-000-A98-133 (ODPRT).  
\bibliographystyle{IEEEtran}
\bibliography{isitbib}

\begin{thebibliography}{10}
\providecommand{\url}[1]{#1}
\csname url@samestyle\endcsname
\providecommand{\newblock}{\relax}
\providecommand{\bibinfo}[2]{#2}
\providecommand{\BIBentrySTDinterwordspacing}{\spaceskip=0pt\relax}
\providecommand{\BIBentryALTinterwordstretchfactor}{4}
\providecommand{\BIBentryALTinterwordspacing}{\spaceskip=\fontdimen2\font plus
\BIBentryALTinterwordstretchfactor\fontdimen3\font minus
  \fontdimen4\font\relax}
\providecommand{\BIBforeignlanguage}[2]{{%
\expandafter\ifx\csname l@#1\endcsname\relax
\typeout{** WARNING: IEEEtran.bst: No hyphenation pattern has been}%
\typeout{** loaded for the language `#1'. Using the pattern for}%
\typeout{** the default language instead.}%
\else
\language=\csname l@#1\endcsname
\fi
#2}}
\providecommand{\BIBdecl}{\relax}
\BIBdecl

\bibitem{GP80}
S.~Gelfand and M.~Pinsker, ``{Coding for channel with random parameters},''
  \emph{Prob. of Control and Inf. Th.}, vol.~9, no.~1, pp. 19--31, 1980.

\bibitem{elgamal}
A.~{El~Gamal} and Y.-H. Kim, \emph{Network Information Theory}.\hskip 1em plus
  0.5em minus 0.4em\relax Cambridge, U.K.: Cambridge University Press, 2012.

\bibitem{tyagi}
H.~Tyagi and P.~Narayan, ``The {Gelfand-Pinsker} channel: Strong converse and
  upper bound for the reliability function,'' in \emph{Proc. of IEEE Intl.\
  Symp.\ on Info.\ Theory}, Seoul, Korea, 2009.

\bibitem{Csi97}
I.~Csisz\'{a}r and J.~{K\"{o}rner}, \emph{Information Theory: Coding Theorems
  for Discrete Memoryless Systems}.\hskip 1em plus 0.5em minus 0.4em\relax
  Cambridge University Press, 2011.

\bibitem{Mou03}
P.~Moulin and J.~A. O'Sullivan, ``Information-theoretic analysis of information
  hiding,'' \emph{IEEE Trans. on Inf. Th.}, vol.~49, no.~3, pp. 563–--593, Mar
  2003.

\bibitem{VH94}
S.~Verd\'{u} and T.~S. Han, ``A general formula for channel capacity,''
  \emph{IEEE Trans. on Inf. Th.}, vol.~40, no.~4, pp. 1147--57, Apr 1994.

\bibitem{Han10}
T.~S. Han, \emph{Information-Spectrum Methods in Information Theory}.\hskip 1em
  plus 0.5em minus 0.4em\relax Springer Berlin Heidelberg, Feb 2003.

\bibitem{iwata02}
K.-I. Iwata and J.~Muramatsu, ``An information-spectrum approach to
  rate-distortion function with side information,'' \emph{IEICE Trans.\ on
  Fundamentals of Electronics, Communications and Computer}, vol. E85-A, no.~6,
  pp. 1387--95, 2002.

\bibitem{Wyner75}
A.~D. Wyner, ``On source coding with side information at the decoder,''
  \emph{IEEE Trans. on Inf. Th.}, vol.~21, no.~3, pp. 294--300, 1975.

\bibitem{Ste08}
Y.~Steinberg, ``Coding for channels with rate-limited side information at the
  decoder, with applications,'' \emph{IEEE Trans. on Inf. Th.}, vol.~54, no.~9,
  pp. 4283--95, Sep 2008.

\bibitem{heegard}
C.~Heegard and A.~{El Gamal}, ``On the capacity of computer memory with
  defects,'' \emph{IEEE Trans. on Inf. Th.}, vol.~29, no.~5, pp. 731--739, May
  1983.

\bibitem{Ste96}
Y.~Steinberg and S.~Verd\'{u}, ``Simulation of random processes and
  rate-distortion theory,'' \emph{IEEE Trans. on Inf. Th.}, vol.~42, no.~1, pp.
  63--86, Jan 1996.

\bibitem{mat12}
T.~Matsuta and T.~Uyematsu, ``A general formula of rate-distortion functions
  for source coding with side information at many decoders,'' in \emph{Int.
  Symp. Inf. Th.}, Boston, MA, 2012.

\bibitem{miyake}
S.~Miyake and F.~Kanaya, ``Coding theorems on correlated general sources,''
  \emph{IEICE Trans.\ on Fundamentals of Electronics, Communications and
  Computer}, vol. E78-A, no.~9, pp. 1063--70, 1995.

\bibitem{Hayashi06}
M.~Hayashi, ``General nonasymptotic and asymptotic formulas in channel
  resolvability and identification capacity and their application to the
  wiretap channel,'' \emph{IEEE Trans. on Inf. Th.}, vol.~52, no.~4, pp.
  1562--1575, April 2006.

\bibitem{bloch13}
M.~Bloch and J.~N. Laneman, ``Strong secrecy from channel resolvability,''
  \emph{IEEE Trans. on Inf. Th.}, vol.~59, no.~12, pp. 8077--8098, Dec 2013.

\bibitem{WeiYu}
W.~Yu, A.~Sutivong, D.~Julian, T.~Cover, and M.~Chiang, ``Writing on colored
  paper,'' in \emph{Int. Symp. Inf. Th.}, {Washington, DC}, 2001.

\bibitem{Mou07}
P.~Moulin and Y.~Wang, ``Capacity and random-coding exponents for channel
  coding with side information,'' \emph{IEEE Trans. on Inf. Th.}, vol.~53,
  no.~4, pp. 1326--47, Apr 2007.

\bibitem{Kuz12}
S.~Kuzuoka, ``A simple technique for bounding the redundancy of source coding
  with side information,'' in \emph{Int. Symp. Inf. Th.}, Boston, MA, 2012.

\bibitem{coverChiang}
T.~M. Cover and M.~Chiang, ``Duality between channel capacity and rate
  distortion with two-sided state information,'' \emph{IEEE Trans. on Inf.
  Th.}, vol.~48, no.~6, pp. 1629--38, Jun 2002.

\bibitem{HV93}
T.~S. Han and S.~Verd\'{u}, ``Approximation theory of output statistics,''
  \emph{IEEE Trans. on Inf. Th.}, vol.~39, no.~3, pp. 752--72, Mar 1993.

\bibitem{rag13}
M.~Raginsky, ``Empirical processes, typical sequences and coordinated actions
  in standard {Borel} spaces,'' \emph{IEEE Trans. on Inf. Th.}, vol.~59, no.~3,
  pp. 1288--1301, Mar 2013.

\bibitem{Rosen05}
A.~Rosenzweig, Y.~Steinberg, and S.~Shamai, ``On channels with partial channel
  state information at the transmitter,'' \emph{IEEE Trans. on Inf. Th.},
  vol.~51, no.~5, pp. 1817--30, 2005.

\bibitem{verdu12}
S.~Verd\'u, ``Non-asymptotic achievability bounds in multiuser information
  theory,'' in \emph{Allerton Conference}, 2012.

\bibitem{WKT13}
S.~Watanabe, S.~Kuzuoka, and V.~Y.~F. Tan, ``Non-asymptotic and second-order
  achievability bounds for coding with side-information,''
  \emph{arXiv:1301.6467}, Jan 2013.

\bibitem{YAG13b}
M.~H. Yassaee, M.~R. Aref, and A.~Gohari, ``A technique for deriving one-shot
  achievability results in network information theory,''
  \emph{arXiv:1303.0696}, Mar 2013.

\end{thebibliography}

\begin{IEEEbiography}[{\includegraphics[width=1in,height=1.25in,clip,keepaspectratio]{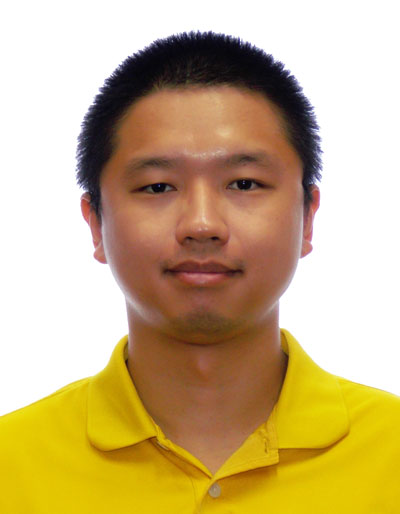}}]{Vincent Y. F. Tan} (S'07-M'11)  is an Assistant Professor   in the Departments of Electrical and Computer
Engineering (ECE) and Mathematics at the National
University of Singapore.  He received the B.A.\ and M.Eng.\ degrees in Electrical and Information Sciences from  Cambridge
University in 2005. He received the Ph.D.\ degree in Electrical Engineering and
Computer Science  from the Massachusetts Institute of Technology in 2011. He was
  a postdoctoral researcher in the Department of  ECE  at the University of Wisconsin-Madison and following
that, a scientist at the Institute for Infocomm (I$^2$R) Research,  A*STAR, Singapore. His research interests include information theory as well as  learning and inference of graphical models.

Dr.\ Tan   received the 2011 MIT EECS Jin-Au Kong outstanding doctoral thesis prize. He is a member of the
IEEE Machine Learning for Signal Processing (MLSP) Technical Committee and a Technical Program Committee member of the 2014 International Symposium on Information Theory. 
\end{IEEEbiography}
\end{document}